%% file: main.tex
\documentclass[twocolumn]{aastex631}
\usepackage{amsmath, amssymb, amsthm, makecell}
\usepackage{enumitem}
\usepackage{multirow}
\usepackage{tabularx}
\usepackage{hyperref}
\usepackage[hang]{footmisc}

\newcommand{\ftnote}[1]{\footnote{
    \hspace{-10pt}
    \vtop{\hsize=\dimexpr\linewidth+10pt\relax 
    #1}
}}
\renewcommand{\la}{\langle}
\newcommand{\ra}{\rangle}
\newcommand{\CMB}{\textrm{CMB}}
\newcommand{\fg}{\textrm{fg}}
\newcommand{\alens}{A_{\textrm{lens}}}

\renewcommand{\O}{\mathcal{O}}

\usepackage{soul}

\begin{document}

    \title{The Simons Observatory: Assessing the Impact of Dust Complexity on the Recovery of Primordial $B$-modes}
    
    \input{author_list}

    \begin{abstract}
        We investigate how dust foreground complexity can affect measurements of the tensor-to-scalar ratio, $r$, in the context of the Simons Observatory, using a cross-spectrum component separation analysis. 
        Employing a suite of simulations with realistic Galactic dust emission, 
        we find that spatial variation in the dust frequency spectrum, parametrized by
        $\beta_d$, can bias the estimate for $r$ when modeled using a 
        low-order moment expansion to capture this spatial variation. While this approach performs well across 
        a broad range of dust complexity, the bias increases with more extreme spatial 
        variation in dust frequency spectrum, reaching as high 
        as $r\sim0.03$ for simulations with no primordial tensors and a spatial dispersion of 
        $\sigma(\beta_d)\simeq0.3$---the most extreme case considered, yet still consistent with current observational constraints.
        This bias is driven by changes in the $\ell$-dependence of the dust power spectrum as a function of frequency that can mimic a primordial $B$-mode tensor signal. Although low-order moment expansions fail to capture the full effect when the spatial variations of $\beta_d$ become large and highly non-Gaussian, our results show that extended parametric methods can still recover unbiased estimates of $r$ under a wide range of dust complexities. We further find that the bias in $r$, at the highest degrees of dust complexity, is largely insensitive to the spatial structure of the dust amplitude and is instead dominated by spatial correlations between $\beta_d$ and dust amplitude, particularly at higher orders. 
        If $\beta_d$ does spatially vary at the highest levels investigated here, we would expect to use more flexible foreground models to achieve an unbiased constraint on $r$ for the noise levels anticipated from the Simons Observatory.
    \end{abstract}
    
    \keywords{Cosmology (343), Cosmic inflation (319), CMB (322), Diffuse radiation (383)}
        
    \section{Introduction}
    \label{sec:intro}
    
    A detection of primordial gravitational waves (PGWs) through their imprint on the polarization of the cosmic microwave background (CMB) would offer a powerful probe of the physics of the early Universe. PGWs arise as tensor perturbations of the metric and generate a parity-odd polarization pattern in the CMB, known as the primordial $B$-mode, via Thomson scattering of anisotropic radiation off free electrons at recombination and reionization \citep{kamionkowski_1997_primodial_gw, seljak_1997_gw_polarization}. The amplitude of the primordial tensor perturbations compared to the primordial density perturbations is parametrized as the tensor-to-scalar ratio, $r$. Tensor perturbations are a generic prediction of inflationary models, produced during a period of rapid expansion in the early Universe \citep{guth_1981_inflation, linde_1982_inflation}, and their amplitude directly relates to the energy scale of inflation.  Constraints from the joint analysis of BICEP/Keck, WMAP, and Planck data limit $r < 0.032-0.036$ at the 95\% confidence level \citep{bicepkeck_2021_r_constraint,Tristram_2022_bicep_npipe_r}.

    Targeting this signal has motivated the design of a new generation of CMB polarization experiments, 
    both ground-based and space-borne, that aim to survey the polarized microwave sky with unprecedented 
    sensitivity and frequency coverage over the coming decade. Several of these, such as the BICEP/Keck Array, 
    the Simons Observatory (SO), and CMB-S4, will focus on detecting degree-scale $B$-modes associated 
    with the ``recombination bump" in the CMB power spectrum around $\ell \sim 100$ \citep{SO_2019_forecast,
    Ade_2022_BICEP3_design, Abazajian_2022_CMBS4_r_forecast}. While ground-based observations are limited by
    atmospheric noise on the largest angular scales, they will be highly sensitive at intermediate scales. 
    In parallel, the future \textit{LiteBIRD} satellite mission is expected to target both the recombination and 
    reionization bumps (the latter peaking at $\ell < 10$), with greater sky and frequency coverage 
    \citep{LiteBIRD_2023_forecast}. Together, these efforts are expected to improve current constraints 
    on $r$ by more than an order of magnitude.
    
    Achieving this level of precision, however, requires overcoming a range of observational challenges. 
    From the sky, the two dominant sources of contamination to the primordial $B$-mode signal are 
    lensing-induced $B$-modes and Galactic foregrounds. Gravitational lensing by large-scale structure 
    deflects CMB photons, converting part of the primordial $E$-mode polarization into $B$-modes, 
    particularly at smaller angular scales ($\ell\gtrsim 100$) 
    \citep{Zaldarriaga_1998_lensing_bmode, lewis_2006_weak_lensing}. 
    Galactic emission dominates over the primordial $B$-mode signal at all frequencies and angular scales: 
    synchrotron radiation from relativistic electrons in magnetic fields is strongest at lower frequencies 
    ($\lesssim$ 60\,GHz), while thermal emission from aligned interstellar grains becomes dominant at higher 
    frequencies ($\gtrsim$ 100\,GHz) \citep{planck_2020_dust}. 
    
    If the Galactic foreground emission exhibited a uniform spectral energy distribution (SED) across the sky (i.e., perfect correlation between frequencies), the removal of these $B$-mode foregrounds would be straightforward. One could measure each component at its dominant frequency (low for synchrotron and high for dust) and then simply scale the map to other frequencies using a single scalar parameter. However, spatial variation in the frequency dependence of the foreground SEDs results in ``frequency decorrelation,'' meaning that maps at different frequencies cannot be related by a simple multiplicative factor \citep{Planck_2016_XXX, Planck_2017_L_spatial_vary_SED}. In addition, the Galactic magnetic field structure can produce
    line-of-sight (LOS) variation in SED parameters, creating a frequency-dependent polarization angle 
    known as ``LOS decorrelation'' \citep{tassis, Vacher2023}. When spectral parameters governing SEDs 
    vary across the sky, computing angular power spectra from maps involves averaging over regions 
    with different SEDs. Due to the non-linearity of the SED variation, the averaging leads to a 
    phenomenon known as ``SED distortion'' such that even if every SED on every line-of-sight could 
    be described by a simple parametric model, the averaged SED may not be able to be fit by the same 
    model \citep{Chluba_2017_LOS}.
      
    For current experiments, these effects are particularly important for thermal dust emission, 
    which may exhibit complex spectral behavior due to physical variations in both the dust temperature 
    ($T_d$) and spectral index ($\beta_d$) throughout the interstellar medium. These variations can 
    occur both spatially on the sky and along the LOS, as different dust populations with varying 
    properties contribute to the observed signal. 
    Synchrotron emission can also exhibit frequency decorrelation, primarily driven by spatial variations in its spectral index, although its contribution is typically subdominant at the frequencies targeted by upcoming experiments focused on detecting primordial $B$-modes.

    The level of frequency decorrelation in polarized dust and synchrotron emission at millimeter 
    wavelengths remains uncertain. Planck data have constrained a 97.5\% dust correlation
    limit between the 217 and 353~GHz channels at 99.1\% across scales $50 < \ell < 160$ over 71\% of 
    the sky \citep{planck_2020_dust}. 
    These global constraints do not rule out more significant decorrelation in smaller, localized regions. 
    More targeted studies have probed line-of-sight decorrelation, revealing small-scale spatial variations in the decorrelation degree of the dust  \citep{pelgrims2021, Ritacco:2023}. 
    Complementary results from BICEP/Keck find no significant decorrelation in clean, high-latitude 
    regions relevant for primordial $B$-mode searches \citep{2023ApJ...945...72A}, whereas the SPIDER 
    balloon experiment reports evidence for spatial variation in dust spectral behavior in similar 
    regions \citep{Ade:2025}. Importantly, viable models of dust polarization allow for decorrelation 
    levels that can vary by orders of magnitude between 150~GHz and higher frequencies 
    \citep{planck_2020_dust, ThePan-ExperimentGalacticScienceGroup:2025}.

    If unmodeled, decorrelation from spatial and line-of-sight variations in foregrounds spectral properties can bias component separation and lead to incorrect inferences of primordial $B$-modes \citep{Remazeilles_2016_r_bias_fg_model, Planck_2017_L_spatial_vary_SED}. This effect is expected to be increasingly important for high-sensitivity experiments including the SO. The SO, a ground-based experiment located in the Atacama Desert, will observe the microwave sky in six frequency bands spanning 27–280~GHz. Initial observations are underway. Its Small Aperture Telescopes (SATs) are designed to target large-scale polarization over $\sim10-20$\% of the sky, aiming to constrain $\sigma(r)$ to an uncertainty of 0.002--0.003  \citep{SO_2019_forecast}. At this sensitivity level, characterizing and mitigating possible biases arising from mismodeling the effects of frequency decorrelation is important for robust component separation and  cosmological inference.

    A commonly used approach in current and past CMB analyses is the cross-spectrum ($C_\ell$-based) 
    component separation method, where auto- and cross-power spectra between frequency 
    channels—particularly in $B$-modes—are used to construct a data vector that is fit with a 
    model of cosmological and Galactic components \citep[e.g.,][]{bicep_2016_r_constraint, 
    bicepkeck_2021_r_constraint, Planck_2016_dr2_likelihood, Planck_2018_cosmo_params}. 
    This method is especially well suited to ground-based experiments, where complex time-stream filtering 
    and spatially inhomogeneous, non-white noise introduce non-trivial correlations between 
    pixels, making map-based likelihoods challenging to implement reliably \citep[e.g.,][]{Hervias_2025_so_tf}.
    In its simplest form, the $C_\ell$-based approach assumes perfect frequency correlation of foregrounds, i.e., that the foreground signal at a given frequency can be extrapolated from a template at a pivot frequency with a single parametrized spectral scaling. 
    
    To relax this assumption, analytic extensions based on moment expansions have been developed to model 
    frequency decorrelation more flexibly \citep{Chluba_2017_LOS, azzoni_2021_bbpower_moment, 
    Remazeilles_2021_fg_cmILC, Mangilli2021, Vacher_2022_dust_moment, 
    azzoni_2023_hybrid_bbpower, Carones_2024_moment_model}.
    While these extensions have shown success in mitigating possible biases in the inferred cosmological parameters, it remains relatively unexplored which physical aspects of foreground complexity — such as spatial variation in the spectral parameters, LOS mixing, 
    correlations between spectral parameters and foreground amplitudes — are primarily responsible for producing these biases. 
    Recent work by \citet{Vacher_2024_how_bad_dust} investigated aspects of this question in the context of the \textit{LiteBIRD} mission, using map-based component separation, with a spin-moment expansion to model LOS dust variations, and showed that unmodeled moment complexity can bias $r$, reinforcing the need for physically motivated foreground models.
    Given the current uncertainties in the level and origin of frequency decorrelation, and the growing use of increasingly complex yet observationally plausible foreground models, it is important to identify which aspects of this complexity impact cosmological inference the most, and to assess the robustness of existing mitigation strategies in such regimes. 

    In this work, we build on the spectrum-based $B$-mode analysis described in \cite{Wolz_2023_pipeline_comparison} (referred to there as ``pipeline A'') for application to SO, working towards quantifying the impact of frequency decorrelation on the measurement of $r$ using state-of-the-art foreground simulations. By using physically parametrized models, we separately study the impact of spatial variations in the dust temperature, $T_d$, and the dust spectral index, $\beta_d$, on estimates for $r$. 
    We also approximately assess the likely impact of line-of-sight (LOS) frequency decorrelation of the foregrounds, which would result in a frequency-dependent polarization angle, on estimates of $r$. To achieve this goal, we simulate different foreground realizations to assess how foreground complexity affects 
    the $C_\ell$-based component separation method. In this paper, Section~\ref{sec:sims} describes the simulations for the analysis, Section~\ref{sec:cross_cl} summarizes the cross $C_\ell$ foreground cleaning formalism, Section~\ref{sec:result} analyzes the effects of foreground complexity on estimates of $r$, Section~\ref{sec:discussion} presents a discussion of the results, and Section~$\ref{sec:conclusion}$ concludes with a summary.

    \input{simulations}
    
    \input{pipeline}

    \input{result_section}

    \input{discussion}

    \input{conclusion}

    \section*{Acknowledgements}
    The authors would like to thank Irene Abril-Cabezas for useful feedback.
    This work was supported in part by a grant from the Simons Foundation (Award \#457687, B.K.).
    The simulations presented in this article were performed on computational resources managed and supported by Princeton Research Computing.
    This work was carried out in part at the Jet Propulsion Laboratory, California Institute of Technology, under a contract with the National Aeronautics and Space Administration.
    DA acknowledges support from STFC under grant ST/W000903/1, and from the Beecroft Trust.
    CB acknowledges partial support by the Italian Space Agency \textit{LiteBIRD} Project (ASI Grants No. 2020-9-HH.0 and 2016-24-H.1-2018), as well as the InDark and \textit{LiteBIRD} Initiative of the National Institute for Nuclear Physics, and the RadioForegroundsPlus Project HORIZON-CL4-2023-SPACE-01, GA 101135036, and Project SPACE-IT-UP by the Italian Space Agency and Ministry of University and Research, Contract Number 2024-5-E.0.
    MLB acknowledges support from the UKRI/STFC (grant number ST/X006344/1).
    SEC acknowledges support from NSF grant No. AST-2441452, and from an Alfred P. Sloan Research Fellowship. 
    JD acknowledges support from a Royal Society Wolfson Visiting Fellowship and from the Kavli Institute for Cosmology Cambridge and the Institute of Astronomy, Cambridge.
    NK acknowledges partial support from the InDark Initiative of the National Institute for Nuclear Phyiscs (INFN), and the RadioForegroundsPlus Project HORIZON-CL4-2023-SPACE-01, GA 101135036, and Project SPACE-IT-UP by the Italian Space Agency and Ministry of University and Research, Contract Number 2024-5-E.0.
    MR acknowledges the support of the Spanish Ministry of Science and Innovation through grants PID2022-139223OB-C21 and PID2022-140670NA-I00, as well as the RadioForegroundsPlus Project HORIZON-CL4-2023-SPACE-01, GA 101135036.
    LV acknowledges partial support from the RadioForegroundsPlus Project HORIZON-CL4-2023-SPACE-01, GA 101135036.
    We acknowledge the use of \texttt{CAMB} \citep{Lewis_2002_camb}, \texttt{emcee} \citep{Foreman_mackey_2013_emcee}, \texttt{healpy} \citep{Zonca_2019_healpy}, \texttt{matplotlib} \citep{hunter_2007_matplotlib}, \texttt{numpy} \citep{harris_2020_numpy}, \texttt{PySM3} \citep{Thorne_2017_pysm, Zonca2021, ThePan-ExperimentGalacticScienceGroup:2025}, and \texttt{scipy} \citep{Virtanen_2020_SciPy} software packages.

    \bibliography{references}
    \bibliographystyle{aasjournal}
\end{document}

%% file: author_list.tex
\author[0000-0002-5210-8035]{Yiqi Liu}
\affiliation{Department of Physics, Princeton University, Jadwin Hall, Princeton, NJ 08544, USA}

\author[0000-0002-8132-4896]{Susanna Azzoni}
\affiliation{Department of Physics, Princeton University, Jadwin Hall, Princeton, NJ 08544, USA}

\author[0000-0002-7633-3376]{Susan E. Clark}
\affiliation{Department of Physics, Stanford University, Stanford, CA 94305, USA} 
\affiliation{Kavli Institute for Particle Astrophysics \& Cosmology, P.O. Box 2450, Stanford University, Stanford, CA 94305, USA}

\author[0000-0001-7449-4638]{Brandon S. Hensley}
\affiliation{Jet Propulsion Laboratory, California Institute of Technology, 4800 Oak Grove Drive, Pasadena, CA 91109, USA}

\author[0000-0001-9551-1417]{L\'eo Vacher}
\affiliation{The International School for Advanced Studies (SISSA), via Bonomea 265, I-34136 Trieste, Italy}

\author[0000-0002-4598-9719]{David Alonso}
\affiliation{Department of Physics, University of Oxford, Denys Wilkinson Building, Keble Road, Oxford OX1 3RH, United Kingdom}

\author[0000-0002-8211-1630]{Carlo Baccigalupi}
\affiliation{The International School for Advanced Studies (SISSA), via Bonomea 265, I-34136 Trieste, Italy}
\affiliation{The National Institute for Nuclear Physics (INFN), via Valerio 2, I-34127, Trieste, Italy} \affiliation{The Institute for Fundamental Physics of the Universe (IFPU), Via Beirut 2, I-34151, Trieste, Italy}

\author[0000-0002-0370-8077]{Michael L. Brown}
\affiliation{Jodrell Bank Centre for Astrophysics, Department of Physics and Astronomy, Alan Turing Building, The University of Manchester, Manchester M13 9PL, United Kingdom}

\author[0009-0004-1436-841X]{Alessandro Carones}
\affiliation{The International School for Advanced Studies (SISSA), via Bonomea 265, I-34136 Trieste, Italy}

\author[0000-0003-3725-6096]{Jens Chluba}
\affiliation{Jodrell Bank Centre for Astrophysics, Department of Physics and Astronomy, Alan Turing Building, The University of Manchester, Manchester M13 9PL, United Kingdom}

\author[0000-0002-7450-2586]{Jo Dunkley}
\affiliation{Department of Physics, Princeton University, Jadwin Hall, Princeton, NJ 08544, USA}
\affiliation{Department of Astrophysical Sciences, Princeton University, Peyton Hall, Princeton, NJ 08544, USA}

\author[0000-0002-4765-3426]{Carlos Herv\'ias-Caimapo}
\affiliation{Instituto de Astrof\'isica and Centro de Astro-Ingenier\'ia, Facultad de F\'isica, Pontificia Universidad Cat\'olica de Chile, Av. Vicu\~na Mackenna 4860, 7820436 Macul, Santiago, Chile}

\author[0000-0002-6898-8938]{Bradley R. Johnson}
\affiliation{Department of Astronomy,  University of Virginia, Charlottesville, VA 22904, USA}

\author[0000-0002-5501-8449]{Nicoletta Krachmalnicoff}
\affiliation{The International School for Advanced Studies (SISSA), via Bonomea 265, I-34136 Trieste, Italy}
\affiliation{The National Institute for Nuclear Physics (INFN), via Valerio 2, I-34127, Trieste, Italy} \affiliation{The Institute for Fundamental Physics of the Universe (IFPU), Via Beirut 2, I-34151, Trieste, Italy}

\author[0000-0002-0689-4290]{Giuseppe Puglisi}
\affiliation{Dipartimento di Fisica e Astronomia, Universit\`a degli Studi di Catania, via S. Sofia, 64, 95123, Catania, Italy}
\affiliation{The National Institute for Nuclear Physics (INFN) - Sezione di Catania, Via S. Sofia 64, 95123 Catania, Italy}
\affiliation{INAF - Osservatorio Astrofisico di Catania, via S. Sofia 78, 95123 Catania, Italy}

\author[0000-0001-9126-6266]{Mathieu Remazeilles}
\affiliation{Instituto de Fisica de Cantabria (CSIC-UC), Avenida de los Castros s/n, 39005 Santander, Spain}

\author[0000-0003-3155-6151]{Kevin Wolz}
\affiliation{Department of Physics, University of Oxford, Denys Wilkinson Building, Keble Road, Oxford OX1 3RH, United Kingdom}

%% file: simulations.tex
    \section{Simulations}
    \label{sec:sims}
    This section describes the simulations we use for the study. We generate simulations using the \textsf{HEALPix}\ftnote{\url{https://healpix.sourceforge.io}} pixelization scheme \citep{Gorski2005} at 
    a resolution of $N_{\textrm{side}} = 512$. 
    This resolution is consistently used for foreground amplitude and SED parameter templates.
    Each realization includes four data splits, each with independent noise, and each split contains six channels corresponding to the SO SAT frequency bands centered at 
    27, 39, 93, 145, 225 and 280~GHz.

    The simulated maps include both sky signal and instrumental noise and are similar to those in \cite{Wolz_2023_pipeline_comparison}, with differences noted where relevant.
    The sky signal comprises the CMB (with input $r = 0$) and Galactic foregrounds, specifically dust and synchrotron emission, with different degrees of complexity, as described below in Section~\ref{sec:sim-fg}. Each signal is convolved with a Gaussian beam with FWHM at each frequency listed in Table~\ref{tab:freq-resol}. To simplify the simulations relative to expected observations, we assume $\delta$-function passbands for each channel.

    \begin{table}[t]
        \centering
        \hspace{-50pt}
        \begin{tabular}{c||c|c|c|c|c|c}
        \hline\hline
            Frequency (GHz)         & 27 & 39 & 93 & 145 & 225 & 280\\
            FWHM (arcmin)           & 91 & 63 & 30 & 17  & 11  & 9  \\
            $N_{\textrm{white}}$ ($\mu$K$\cdot$arcmin)    & 46 & 28 & 3.5& 4.4 & 8.4 & 21 \\
            $\ell_{\textrm{knee}}$  & 15 & 15 & 25 & 25  & 35  & 40 \\
            $\alpha_{N}$            &-2.4&-2.4&-2.5&-3.0 &-3.0 &-3.0\\
            \hline\hline
        \end{tabular}
        \caption{Simulation specifications for the SO SAT frequency channels. Noise parameters correspond to homogeneous white noise levels ($N_{\textrm{white}}$) and $1/f$ noise characterized by a multipole knee ($\ell_{\textrm{knee}}$) and spectral index ($\alpha_{N}$). FWHM values denote the beam size at each frequency. Inhomogeneous noise is modeled by modulating the homogeneous noise with the SO SAT hit-count map, as commonly done in SO simulations \citep[e.g.][]{azzoni_2021_bbpower_moment, Wolz_2023_pipeline_comparison}.}
        \label{tab:freq-resol}
    \end{table}

    \subsection{Noise}
    \label{sec:sim-noise}
    Following \cite{SO_2019_forecast}, we model the noise as a combination of white and red components, with the noise power spectrum given by 
    \begin{align}
        N_\ell = N_{\textrm{white}}\left(1 + \left(\frac{\ell}{\ell_{\textrm{knee}}}\right)^{\alpha_N}\right),
    \end{align}
    where $N_{\textrm{white}}$ is the white noise level, $\ell_{\textrm{knee}}$ and $\alpha_N$ characterize the low-frequency ($1/f$) noise and filtering effects. We adopt the ``baseline optimistic'' noise scenario 
    defined in \citet{SO_2019_forecast}, with specifications for the parameters listed in Table \ref{tab:freq-resol}. 
    As in \cite{Wolz_2023_pipeline_comparison}, homogeneous noise maps are generated from Gaussian 
    realizations of the theoretical $N_\ell$. To produce inhomogeneous noise, we modulate these 
    maps using the SO SAT hit-count map provided in \citet{SO_2019_forecast}.\ftnote{We note that 
    the sky coverage now used in the SO observations has evolved from this forecast version, but
    we do not expect this difference to affect the conclusions of this work.}
    This approach does 
    not fully capture the complexity of the SO noise model, including its anisotropic properties 
    and the impact of filtering on the data \citep{Hervias_2025_so_tf}. However, it provides a 
    reasonable approximation for studying the impact of foreground complexity. Each data split 
    within a simulation uses independent noise realizations.
    
    \subsection{CMB}
    \label{sec:sim-cmb}
    CMB maps are generated as Gaussian realizations of the best-fit $\Lambda$CDM cosmological model from \citet{Planck_2018_cosmo_params}, drawn from a theory spectrum computed using the Python package \textsf{CAMB} \citep{Lewis_2002_camb}.
    The simulations include lensing-induced $B$-modes but no primordial $B$-modes (i.e., $r = 0$). 
   
    We generate 500 CMB realizations for estimating covariance matrices, described in the following section.
    For component separation tests, we produce 100 independent CMB realizations per foreground scenario (see Section~\ref{sec:sim-fg}). 
    
    \subsection{Foregrounds}
    \label{sec:sim-fg}
    Polarized foreground contamination is dominated by Galactic dust and synchrotron emission. We simulate a suite of both Gaussian and more realistic foregrounds: the Gaussian simulations are used for covariance estimation and pipeline validation, while the non-Gaussian simulations are essential for evaluating the impact of spatial and spectral complexity on component separation performance and the robustness of constraints on $r$. We also consider simulations with mixtures of non-Gaussian and Gaussian components. Table~\ref{tab:all_fg_sim_summary} summarizes the simulation scenarios, and we refer to it throughout the following.

    To summarize our simulation design, we categorize the scenarios into three classes:

    The first class includes the \textit{\texttt{d10} Nominal},
    \textit{$\beta_d$-Spatial}, \textit{$T_d$-Spatial}, and \textit{LOS} scenarios. These form the core of our analysis 
    and are used to examine how different aspects of dust complexity can bias the tensor-to-scalar ratio $r$, if the dust is incorrectly modeled. We use the \textit{\texttt{d10} Nominal} scenario to compare results to those obtained
    with the unmodified \texttt{d10} dust foreground.
    
    The second class consists of the \textit{Simple Gauss}, \textit{$\beta_d$ Gauss}, and \textit{$\beta$-Dust Corr.} 
    scenarios. These Gaussian simulations serve both as benchmarks for comparison with previous studies and as the basis
    for covariance construction. Their controlled statistical properties---such as the level of non-Gaussianity and the 
    correlation between $\beta_d$ and dust amplitude---make them useful for isolating the effects of frequency decorrelation 
    on $\chi^2$ values and for exploring the influence of $\beta_d$-dust correlations on possible biases of $r$.
    
    The last class comprises all \textit{Ad10*Bd*} and \textit{Ad10Bd10 rot} scenarios. 
    This class is designed to diagnose the origin of the biases in $r$ that we find in this study. 
    In particular, the \textit{Ad10*Bd*} scenarios construct dust signals with 
    Gaussian and \texttt{d10} dust amplitude and $\beta_d$. 
    The \textit{Ad10Bd10 rot} scenario is designed to investigate the mechanism that we find gives rise to 
    biases in $r$.

    \begin{table*}[t]
        \centering
        \hspace{-80pt}
        \begin{tabular}{c|ccccccccccc}
            \hline\hline
            Scenario & Synch & $\beta_s$ & Dust & $\beta_d$ & $T_d$ & Scaled & Scaling 
            & $N_{\textrm{case}}$ & $N_\textrm{sims}$ & $N_\textrm{cov}$ & Section\\
            \hline
            \texttt{d10} Nominal & \texttt{s5} & \texttt{s5} & \texttt{d10} & \texttt{d10} &
            \texttt{d10} & $\beta_d$ & 1.0 & 1 & 100 & - & \ref{subsec:s5d10-fg}\\
            $\beta_d$-Spatial & \texttt{s5} & \texttt{s5} & \texttt{d10} & \texttt{d10} & 
            $\la\texttt{d10}\ra$ & $\beta_d$ & $[0, 2.];0.2$ & 11 & 1100 & - & \ref{subsec:s5d10-fg}\\
            $T_d$-Spatial & \texttt{s5} & \texttt{s5} & \texttt{d10} & $\la\texttt{d10}\ra$ &
            \texttt{d10} & $T_d$ & $[0, 4.];0.4$ & 11 & 1100 & - & \ref{subsec:s5d10-fg}\\
            LOS & \texttt{s5} & \texttt{s5} & \texttt{d10} 
            & \hspace{-35pt}\makecell{Q: \texttt{d10}\\U: $\la\texttt{d10}\ra$} &
            $\la\texttt{d10}\ra$ & $\beta_d^Q$ & $[0, 2.];0.2$ & 11 & 1100 & - & \ref{subsec:s5d10-fg}\\
            Simple Gauss & PL & -3. & PL & 1.54 & 
            19.6~K & - & - & 1 & 600 & 1 & \ref{subsec:gauss_fg}\\
            $\beta_d$ Gauss & PL & BPL & PL & BPL & 
            19.6~K & $\beta_d$ & $[0, 12.5];2.5$ & 6 & 600 & 6 & \ref{subsec:gauss_fg}\\
            $\beta$-Dust Corr. & PL & BPL & PL & BPL & 
            19.6~K & $\beta_d$ & 12.5 & 1 & 100 & - & \ref{subsec:gauss_fg}\\
            Ad10Bd10 & PL & BPL & \texttt{d10} & \texttt{d10} & 
            19.6~K & $\beta_d$ & 1.6 & 1 & 100 & -  & \ref{subsec:sim-SBd10}\\
            Ad10Bdg & PL & BPL & \texttt{d10} & BPL & 
            19.6~K & $\beta_d$ & 10. & 1 & 100 & -  & \ref{subsec:sim-SBd10}\\
            Ad10gBd10 & PL & BPL & \texttt{d10g} & \texttt{d10} & 
            19.6~K & $\beta_d$ & 1.6 & 1 & 100 & -  & \ref{subsec:sim-SBd10}\\
            Ad10gBdg & PL & BPL & \texttt{d10g} & BPL & 
            19.6~K & $\beta_d$ & 1.6 & 1 & 100 & 1  & \ref{subsec:sim-SBd10}\\
            Ad10Bd10 rots & PL & BPL & \texttt{d10} & BPL & 
            19.6~K & $\beta_d$ & 1.6 & 4. & 400 & - & \ref{subsec:sim-SBd10}\\
            \hline\hline
        \end{tabular}
        \caption{Summary of all simulation scenarios used in this study. The details of each case are described in Section~\ref{sec:sim-fg}. ``Synch'' and ``Dust'' indicate the synchrotron and dust templates, respectively. The template can be a non-Gaussian map (\texttt{s5, d10}), or a Gaussian random realization from a power-law (PL) spectrum or from a power spectrum template (\texttt{d10g}).
        $\beta_s$, $\beta_d$, and $T_d$ are SED parameters, which can be either a constant (a real value or the mean of a template in the SO SAT mask region such as $\la d10\ra$), a realistic template (\texttt{s5, d10}), or 
        a Gaussian random realization from a broken-power-law (BPL) power spectrum. The ``Scaled'' column indicates the SED parameter 
        scaled under the scenario. The ``Scaling'' column indicates the scaling applied to the SED 
        parameters. This scaling is either a constant or a range-increment pair, defined as $x$ 
        in Equation~\ref{eq:decorr_param_scaling}. 
        $N_\textrm{case}$ and $N_\textrm{sims}$ are the number of cases and number of total simulations 
        in each scenario, respectively. If the simulation scenario is used for calculating covariance, $N_\textrm{cov}$
        is the number of covariances constructed from this scenario. A detailed description
        of covariance estimation is presented in Section~\ref{sec:cross_cl-covar}.
        The ``$\beta$-Dust Corr''
        scenario has non-zero $\beta_d$-dust correlation ($\la\beta_d\cdot d\ra\neq 0$), and the 
        ``Ad10Bd10 rots'' scenario applies four different rotations to $\beta_d$ of \texttt{d10}.
        The last column provides the section in which corresponding scenarios are described.}
        \label{tab:all_fg_sim_summary}
    \end{table*}

    \subsubsection{Non-Gaussian Foregrounds}
    \label{subsec:s5d10-fg}
    
    \subsubsection*{Dust} \label{sssec:dust}
    We construct a custom suite of realistic dust models based on the \texttt{d10} model from the \textsf{PySM3} package \citep{Thorne_2017_pysm, Zonca2021, ThePan-ExperimentGalacticScienceGroup:2025}. The \texttt{d10} model is built from the Planck Generalized Needlet Internal Linear Combination (GNILC) dust maps \citep{Remazeilles_2011_GNILC, Planck_2020_compsep}. It uses the Planck PR3 GNILC $Q$/$U$ polarization 
    maps as large-scale spatial templates. 
    SED parameters---dust spectral index $\beta_d$ and temperature $T_d$---are based on fits to the Planck PR2 GNILC-derived maps \citep{Planck_2016_GNILC_dust}.
    The \texttt{d10} model combines these large-scale templates with small-scale fluctuations in both the structure of the polarized dust emission and the parameters of the dust SED.
    The resulting templates are scaled to each observation frequency using a single-component modified blackbody in each pixel, as described in \citet{ThePan-ExperimentGalacticScienceGroup:2025}. 
    
\begin{figure}[t]
            \centering
            \includegraphics[width=\linewidth]{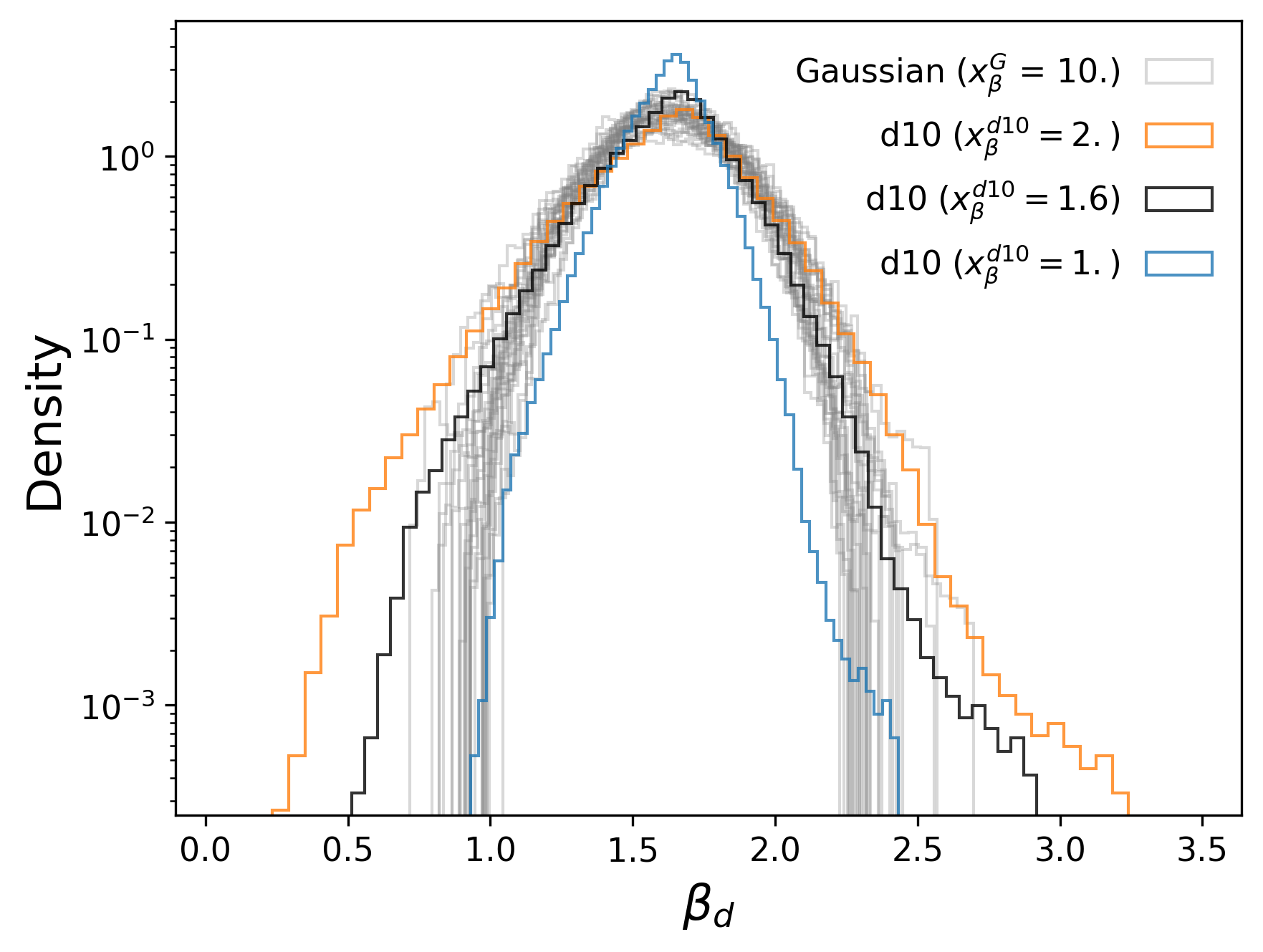}
            \caption{The distribution of dust emissivity indices, $\beta_d$, for the \texttt{d10} simulation in the SO region without
            scaling (blue, $x_{{\beta}}^{{d10}} = 1.$), at the maximum scaling considered
            (orange, $x_{{\beta}}^{{d10}} = 2.$), and at the scaling that matches the degree of decorrelation in the \texttt{d12}  simulation
            (black, $x_{{\beta}}^{{d10}} = 1.6$). These are compared to 30 Gaussian realizations (gray), scaled to $x_{\beta_d}^G=10$, with resulting frequency 
            decorrelation that matches that of the $x_{{\beta}}^{{d10}} = 1.6$ case.} 
            \label{fig:beta_d_hist_max_dispersion}
    \end{figure}

\begin{figure*}[t]
        \centering
        \includegraphics[width=\linewidth]{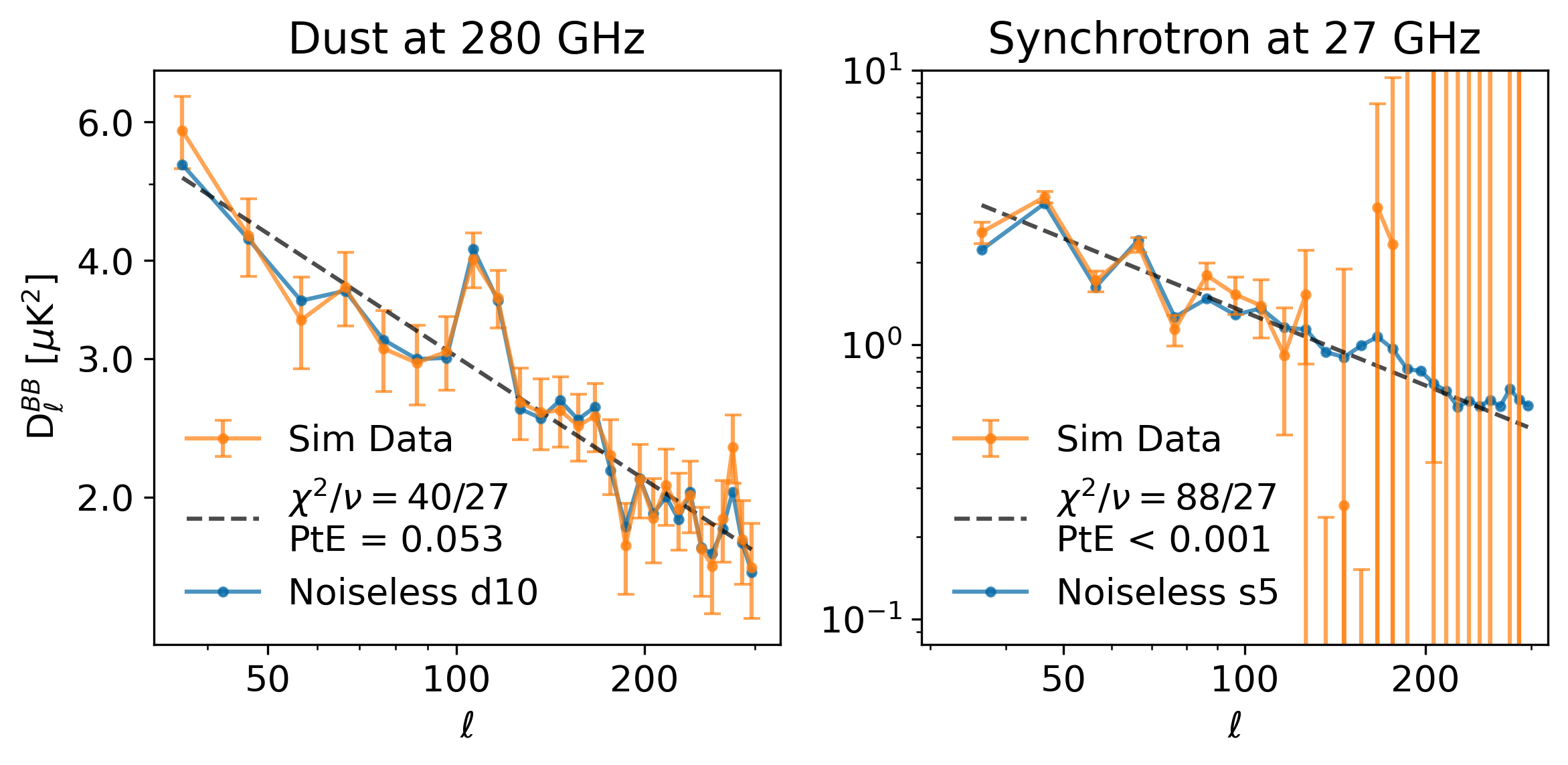}
        \caption{The angular power spectra of the
        \texttt{d10} dust simulation at 280\,GHz (left), and the \texttt{s5} synchrotron simulation at 
        27\,GHz (right), calculated in the SO-SAT region used in this study. 
        In each panel, the orange points shows the binned power spectrum of the simulated foreground maps with noise, and the blue show spectra for noiseless maps.
        The spectra are computed as the mean of cross-split
        spectra, described in Section~\ref{sec:cross_cl}. The errors shown, and their associated covariance matrix, are estimated using the `Cov1' method described in Section \ref{sec:cross_cl-covar}.
        The black dashed line shows the best-fitting power law to the noisy spectra, with the $\chi^2$, degrees of freedom ($\nu$), and 
        probability-to-exceed (PtE) for the fits indicated. The power law is an acceptable fit to the 280~GHz dust spectrum, but a poor fit to the 27~GHz simulated synchrotron spectrum at the larger scales.}
    \label{fig:dust_sync_chi2}
    \end{figure*}

\begin{figure}[ht]
        \centering
        \includegraphics[width=\linewidth]{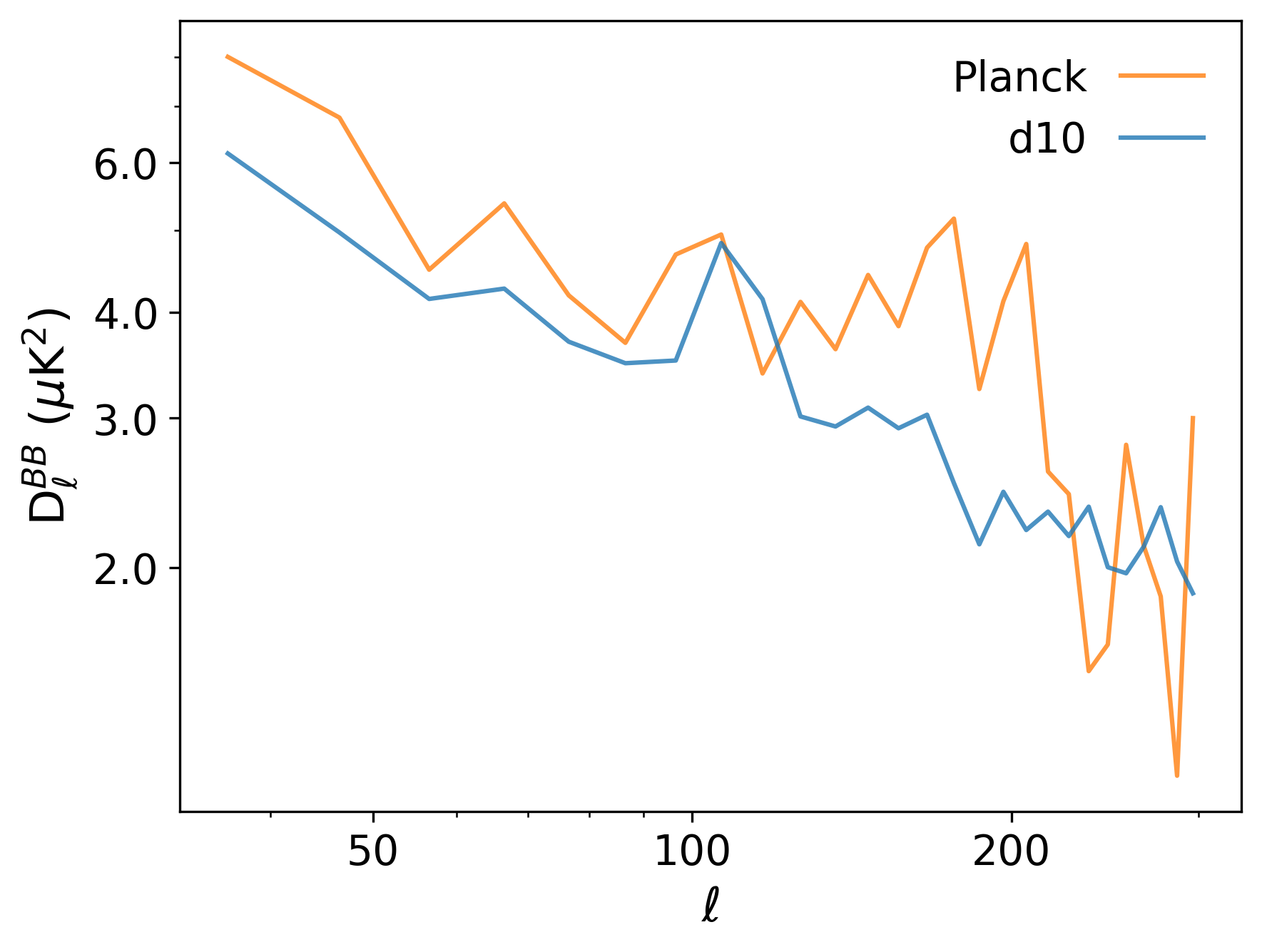}
        \caption{The 217 GHz $\times$ 353 GHz BB cross-spectrum for the \texttt{d10} dust simulation,
        compared to the Planck PR4 data, estimated over the SO-SAT region used in our analysis. At these frequencies the dust signal dominates the Planck spectrum. We note features departing from a pure power-law in both cases.}
        \label{fig:d10_plk}
    \end{figure}

    The distribution of $\beta_d$ values within the sky area considered for SO analysis is shown in Figure~\ref{fig:beta_d_hist_max_dispersion}. As shown in Figure~\ref{fig:dust_sync_chi2}, we find that a power law in $\ell$ provides an acceptable fit to the $BB$ power spectrum of the 
    dust map at 280~GHz, with 
    $\chi^2 = 40$ for 27 dof (PtE of 0.05), where the map includes simulated SO noise described in 
    Section~\ref{sec:sim-noise}. 
    Figure~\ref{fig:d10_plk} shows the $217\times353$~GHz $BB$ cross-spectra for \texttt{d10} dust
    and Planck PR4 measurements \citep{Planck_2020_npipe}. The feature near $\ell \sim 100$ in the \texttt{d10} 
    power spectrum---where it deviates most prominently from a power law---is qualitatively
    consistent with Planck PR4 measurements, which also show significant departures from a pure power law 
    in this multipole range.
    
    In this study we quantify dust complexity using the frequency decorrelation between the
    217~GHz and 353~GHz channels, defined as $1 - \mathcal{R_\ell}^{217\times353}$, where 
    the spectral correlation ratio $\mathcal{R_\ell}^{217\times353}$ is
    \begin{align}
        \label{eq:decorr}
        \mathcal{R}_\ell^{217\times353} = \frac{C_\ell^{217\times353}}{\sqrt{C_\ell^{217\times217}C_\ell^{353\times353}}},
    \end{align}
    as defined in \citet{Planck_2017_L_spatial_vary_SED}.
    We adopt this as our primary proxy for the complexity of a given foreground model, serving as a practical metric to guide our dust model modifications. It is just one of several manifestations of foreground spectral complexity — others include polarization 
    angle rotation, SED distortions, and the frequency dependence of the $EE/BB$ ratio \citep{azzoni_2021_bbpower_moment, Vacher2023_EB}. 
    
    The Planck collaboration placed a 97.5\% limit of $\mathcal{R}_\ell^{217\times353} \geq 0.991$ 
    on the spectral correlation parameter for $50 < \ell < 160$ over the Planck 
    LR71 mask, covering a high-Galactic-latitude region with 71\% effective sky coverage \citep{planck_2020_dust}.
    Uncertainties were quantified based on end-to-end simulations incorporating a model of Galactic magnetic field structure, and are at the $0.005$ level. Among existing \textsf{PySM} models, the \texttt{d12} model based on the ``layer model'' of \citet{Martinez_Solaeche_2018_pysm_d12_mkd} produces decorrelation at a level comparable to this upper limit when evaluated over the LR71 mask \citep{ThePan-ExperimentGalacticScienceGroup:2025}. We therefore use \texttt{d12} as a practical proxy for ``maximal decorrelation'' in our analysis to set the upper bound on spatial variation in the dust SED parameters even though the sky region in this work differs from LR71.

    The SO SAT sky region is smaller than LR71, but the mean decorrelation 
    level of \texttt{d12} over the SO SAT region is only $\sim5\%$ larger, 
    well within the uncertainty estimated on this quantity from the Planck data. 
    The 
    \texttt{d10} model has more decorrelation in the SO SAT region compared to LR71, with $1-\mathcal{R}_\ell^{217\times353} = 
    0.005$ for the SO SAT region we consider, compared to $0.003$ for LR71
    for the $x_{\beta_d} = 1$ case of the 
    \textit{$\beta$-Spatial} scenario.  This difference is within the uncertainty on $\mathcal{R}$ estimated from Planck. Given this sensitivity 
    to sky region and the expectation that the decorrelation level will depend on the exact region chosen for the 
    real SO analysis, we explore a range of decorrelation levels that reach at least the \texttt{d12} level.

    As a reference case, we create the \textit{\texttt{d10} Nominal} scenario, consisting of 100 noise 
    realizations with unmodified \texttt{d10} 
    dust. These serve as a basis for comparing with different \texttt{d10} variants in this study.
    
    To isolate different physical effects, we separately model two sources of decorrelation: \textit{spatial decorrelation}, arising from pixel-to-pixel variation in SED parameters, and \textit{line-of-sight} (LOS) decorrelation, due to the superposition of multiple emission components with different spectra along the same line of sight. These are simulated using distinct approaches described below.

    \begin{figure*}[t]
        \centering
        \includegraphics[width=\linewidth]{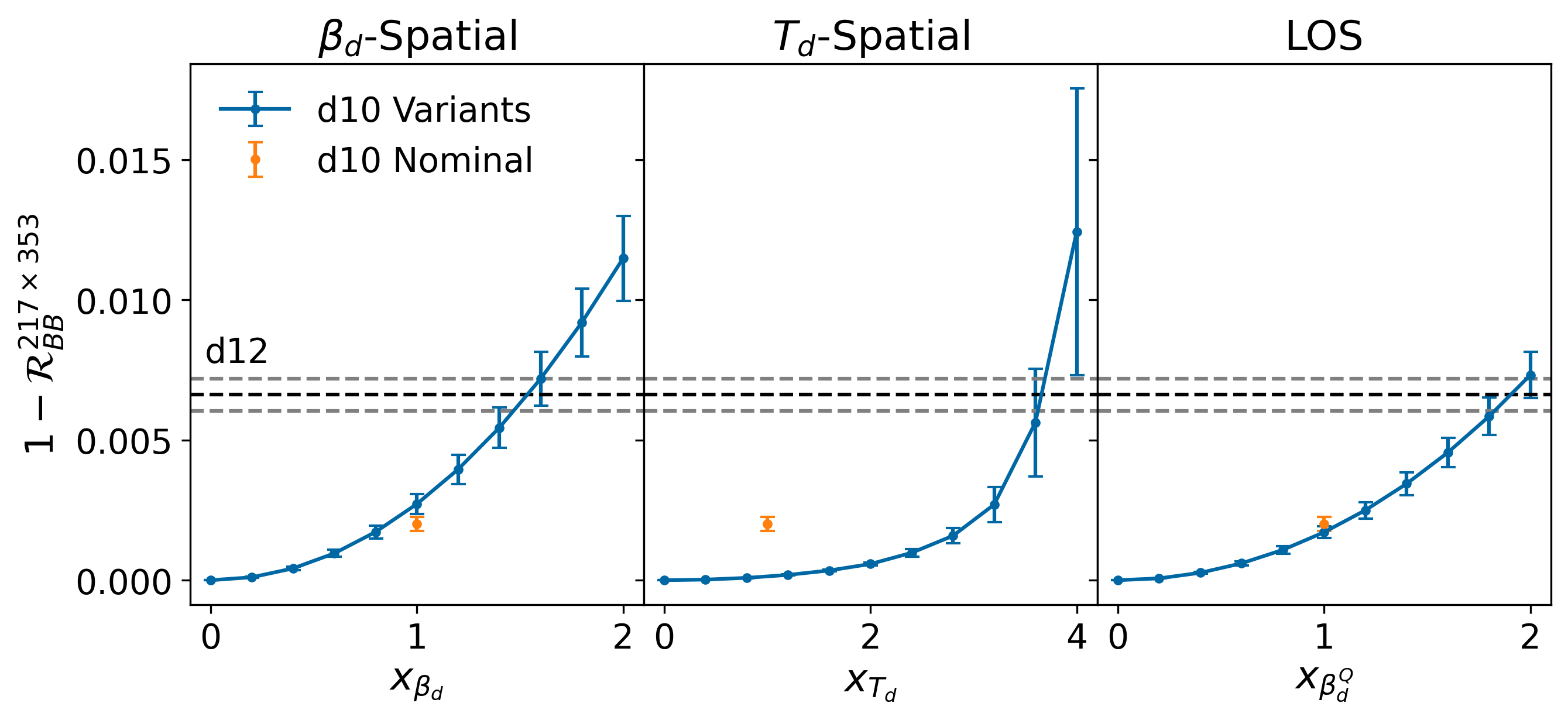}
        \caption{The degree of dust decorrelation ($1 - \mathcal{R}_{BB}^{217\times353}$) between the $217$\,GHz and $353$\,GHz simulated maps, calculated over the Planck LR71 mask and for scales $30 < \ell < 300$, as a function of dust parameter scaling, $x$. 
        The three panels correspond to scaling the $\beta_d$ field (left), the $T_d$ field (middle) and the $\beta_d$ field of just the Q Stokes maps to approximate line-of-sight decorrelation (right). The orange points in each panel show the decorrelation 
        of the \textit{\texttt{d10} Nominal} dust simulation before our modifications.
        The x-axis scalings are specified in Section~\ref{sec:sim-fg}.
        The data points show the decorrelation values averaged over the bandpowers, using the same binning as in Figure~\ref{fig:dust_sync_chi2}, and the errors reflect the dispersion over the bandpowers. The gray dashed 
        line indicates the mean and dispersion for \texttt{d12}.}
        \label{fig:beta_d_scaling}
    \end{figure*}
    \begin{enumerate}[label=$\bullet$]
        \item \textbf{Spatial Decorrelation:} 
        We model spatial decorrelation through spatial variation in the dust SED parameters $\beta_d$ and $T_d$, corresponding to the \textit{$\beta_d$-Spatial} and \textit{$T_d$-Spatial} scenarios listed in Table~\ref{tab:all_fg_sim_summary}. In this case, decorrelation arises from anisotropic maps of SED parameters, where increasing dispersion in either $\beta_d$ or $T_d$ leads to stronger frequency decorrelation.

        To construct these models, we start from the \texttt{d10} SED templates and scale the spatial fluctuations of a chosen SED parameter $\theta$ within the SO SAT patch by a scalar factor $x$, while preserving the mean (monopole) amplitude of the original template:
        \begin{align}
            \theta(\hat{n}, x_\theta) = (\theta(\hat n) - \bar{\theta}) \cdot x_{\theta} + \bar{\theta},
            \label{eq:decorr_param_scaling}
        \end{align}
        where $\theta(\hat n)$ is the unscaled template and $\bar{\theta}$
        its spatial mean within 
        the SAT footprint. $\hat{n}$ determines the template pixel location. 
        The standard deviation of $\theta$
        is then directly proportional to $x$ with this scaling relation. In each simulation, only one parameter ($\beta_d$ or $T_d$) is varied at a time, while the other is held fixed\ftnote{This procedure could, in principle, also be applied to synchrotron parameters such as $\beta_s$, though we focus here on dust-related variations.}.
        For $T_d$, we impose a physical lower bound of 2.725\,K, to avoid unphysical temperatures. 
        
        An important caveat to this procedure arises because the \texttt{d10} templates exhibit a strong anti-correlation between $\beta_d$ and $T_d$. 
        This anti-correlation could be the result of a fitting degeneracy known to impact data of low signal-to-noise ratio \citep{Shetty_2009_dust_sed_param}. However, physical correlations between $\beta_d$ and $T_d$ have been observed in laboratory measurements of interstellar dust analogues \citep{Demyk_2017_dust_sed_params}.  
        This anti-correlation has a counterintuitive effect: increasing the dispersion of one parameter (e.g., $T_d$) while holding the other fixed (e.g., $\beta_d$) at its original template can actually reduce the observed frequency decorrelation between 217 and 353\,GHz. This occurs because the SED scaling 
        anisotropies induced by the two parameters partially cancel out each other when anti-correlated, 
        reducing the net SED scaling anisotropy across the sky and thus lowering the resulting 
        frequency decorrelation.
        
        Since this behavior obscures the intended relationship between SED parameter dispersion and its impact on $r$ bias, we adopt a modified approach in the \textit{$\beta_d$-Spatial} and \textit{$T_d$-Spatial} scenarios (see \ref{tab:all_fg_sim_summary}): when varying one parameter, we fix the other globally to the average value of the original \texttt{d10} template in the SO mask region, rather than using the full spatial map. This allows us to isolate the effect of $\beta_d$ or $T_d$ dispersion independently, avoiding the confounding influence of the strong anti-correlation present in the full \texttt{d10} maps.
        Figure~\ref{fig:beta_d_hist_max_dispersion} shows the distribution of $\beta_d$ within the SO mask region for several scenarios. We compare 30 Gaussian realizations scaled by $x_{\beta_d}^{\rm G} = 10$ (gray, 
        broken power law realizations defined in Section~\ref{subsec:gauss_fg}.) with the unscaled \texttt{d10} distribution (blue), the maximum \texttt{d10} scaling used in this study ($x_{\beta_d}^{\texttt{d10}} = 2.$, orange), and an intermediate scaling ($x_{\beta_d}^{\texttt{d10}} = 1.6$, black), which approximately reproduces the decorrelation level of the \texttt{d12} model. 
        The $x_{\beta_d}^{\rm G}$ defines the scaling of the Gaussian dust $\beta_d$ map, as described in Section~\ref{subsec:gauss_fg}.
        At larger values of $x_{\beta_d}$, the distribution of spectral indices extends beyond the range constrained by current observations 
        in some regions. 

        The relationship between the scaling factor $x$ in Equation~\eqref{eq:decorr_param_scaling} and the resulting frequency decorrelation is shown in Figure~\ref{fig:beta_d_scaling}. We explore $x_{\beta_d} \in [0, 2]$ in steps of 0.2 and $x_{T_d} \in [0, 4]$ in steps of 0.4. At the highest values, the decorrelation reaches $1 - \mathcal{R}_\ell^{217\times353} = 0.012$, surpassing that of \texttt{d12}. 
        These highly decorrelated cases allow us to probe pipeline performance beyond the complexity 
        expected in current data. 
        Notably, achieving a given level of decorrelation requires approximately twice as much 
        scaling in $T_d$ compared to $\beta_d$. This difference likely arises from the 
        spectral behavior of thermal dust emission: with typical dust temperatures around 20~K, 
        the dust SED peaks well above the SO frequency range (approximately 500–700\,GHz). In the 
        Rayleigh-Jeans limit, the dust emission scales linearly with temperature perturbations in 
        each pixel, so $T_d$ 
        variation changes the dust amplitude but not the shape of the SED at first order. As a result, SO 
        observations are relatively insensitive to $T_d$ variations, providing the motivation 
        for neglecting $T_d$ dispersion in \citet{azzoni_2021_bbpower_moment}.

        \item \textbf{Line-of-sight Decorrelation:} 
        This is the \textit{LOS} scenario in Table~\ref{tab:all_fg_sim_summary}.
        Line-of-sight decorrelation can arise when distinct dust populations along the line of sight have different SEDs and polarization angles, leading to the LOS-integrated $Q$ and $U$ Stokes parameters having different SEDs from each other. 
        To simulate this effect in a simplified, artificial manner, we construct dust maps by combining the $Q$ 
        component from simulations with spatially varying $\beta_d$, and the $U$ component from simulations with 
        constant $\beta_d$ ($x_\beta=0$).
        This mismatch in spectral behavior between $Q$ and $U$ mimics the impact of LOS decorrelation. The resulting relation between the decorrelation level and the $\beta_d^Q$ scaling factor is shown in Figure~\ref{fig:beta_d_scaling}. 
        The deliberate suppression of  variation in $\beta_d$ for the 
        $U$ component naturally results in lower frequency decorrelation levels across the cases in 
        the \textit{LOS} scenario compared to those in the \textit{Spatial-$\beta_d$} scenario.
        
        Our implementation captures only one aspect of LOS decorrelation—namely, the spectral rotation of 
        the polarization angle — while neglecting the breakdown of the modified blackbody assumption within 
        individual pixels. Models like \texttt{d12} include more realistic LOS effects, but using them here would 
        introduce confounding factors due to their inherent 2D spatial variations, making it difficult to disentangle
        the LOS contribution. Moreover, the \texttt{d10}-based model allows us to control the level of 
        frequency-decorrelation in a manner consistent with other scenarios used in this study.
        A more systematic approach, such as constructing foreground models from moment maps or small scale filament 
        structure \citep[e.g.,][]{hervias_2022_filament_dust_model, Vacher_2024_how_bad_dust}, 
        is left for future work.
    \end{enumerate}

    \label{subsec:real-foreground}
    \subsubsection*{Synchrotron}
    To model realistic synchrotron emission, we use the \texttt{s5} model from the \textsf{PySM3} Python package \citep{Thorne_2017_pysm, Zonca2021, ThePan-ExperimentGalacticScienceGroup:2025}. This model provides a complex but observationally motivated description of polarized synchrotron at SO frequencies, helping maintain realism in our foreground simulations.
    At the angular scales relevant to this study, \texttt{s5} is formed using synchrotron emission using the Haslam 408~MHz intensity map and WMAP 9-year 23~GHz Q/U polarization maps as spatial templates. The frequency scaling is based on a spatially varying spectral index, $\beta_s$, from \citet{Miville_Deschenes_2008_beta_s} constructed using the Haslam 408~MHz map and the WMAP 3yr K band map
    \citep{Hinshaw_2007_WMAP_3yr, Remazeilles_2015_Haslam} and rescaled to match the level of $\beta_s$ variation seen in a more recent analysis with S-PASS data \citep{Krachmalnicoff_2018_spass_synch}.
    
    Although the primary focus of this study is on the complexity of Galactic dust and its impact on measurements of $r$, we include the \texttt{s5} synchrotron model in our realistic foreground simulations to preserve fidelity to plausible sky conditions. We choose not to vary the synchrotron model across simulations to isolate the effect of dust complexity. 
    
    As shown in Figure~\ref{fig:dust_sync_chi2}, we find that a power law is a 
    poor fit to the  synchrotron power spectrum at 27~GHz, especially at $\ell<80$, with $\chi^2 = 88$ for 27 dof (PtE $< 0.0005$)\ftnote{This statistic is computed using a Gaussian covariance estimated from idealized simulations, which may not fully capture the complexity of the foregrounds.
    This could lead to a mismatch between the assumed uncertainties and the true variation in the data, potentially inflating the $\chi^2$.}.

    \subsubsection{Gaussian Foregrounds}
    \label{subsec:gauss_fg}

    \begin{table*}[ht]
        \centering
        \hspace{-50pt}
        \begin{tabular}{c||cccccccc}
            \hline\hline
            & $A_a^{EE}$ ($\mu K_{\textrm{CMB}}^2$) & $A_a^{BB}$ ($\mu K_{\textrm{CMB}}^2$) & $\alpha_a^{EE}$ 
            & $\alpha_a^{BB}$ & $\beta_a$ & $T_a$ (K)\\
            \hline
            Dust & 56. & 28. & -0.32 & -0.16 & 1.54 & 19.6 \\
            Synchrotron & 9. & 1.6 & -0.7 & -0.93 & -3 & - \\
            \hline\hline
        \end{tabular}
        \caption{Summary of parameters used for Gaussian foregrounds. The angular power spectra parameters are defined in 
        Equation~\eqref{eq:fg_ps}. The SED parameters are defined in Equation~\eqref{eq:dust_sed} and Equation~\eqref{eq:sync_sed} for 
        dust and synchrotron, respectively.}
        \label{tab:gauss-fg-params}
    \end{table*}
    
    We generate Gaussian foreground realizations for both synchrotron and dust by 
    sampling from the power-law angular power spectrum model
    \begin{align}
        \label{eq:fg_ps}
        D_\ell = \frac{\ell(\ell + 1)}{2\pi}C_\ell = A_c^{CC}\left(\frac{\ell}{\ell_0}\right)^{\alpha_c^{CC}},
    \end{align}
    where $c\in\{d, s\}$ denotes dust and 
    synchrotron respectively, $CC\in\{EE, BB\}$ denotes the polarization, $A_c$ is the amplitude and $\alpha_c$ is the spectral index. We fix the pivot scale to $\ell_0 = 80$.
    These realizations are generated at the pivot frequencies and scaled to other frequencies using 
    the SED models. In the Rayleigh-Jeans temperature units, the synchrotron 
    SED follows a power law, and the dust SED follows the modified blackbody:   
    \begin{align}
        \label{eq:dust_sed}
        S_{\nu_0^d}^{d, \nu} &= \left(\frac{\nu}{\nu_0^d}\right)^{\beta_d}\frac{B(T_d, \nu)}{B(T_d, \nu_0^d)},\\
        \label{eq:sync_sed}
        S_{\nu_0^s}^{s, \nu} &= \left(\frac{\nu}{\nu_0^s}\right)^{\beta_s},
    \end{align}
    where $B(T_d, \nu)$ is the blackbody function and we fix the dust temperature to $T_d = 19.6$~K.
    The resulting maps are then converted to CMB temperature units.
    The Gaussian foreground parameters are listed in Table~\ref{tab:gauss-fg-params}. 

    We also generate a variant, referred to later with the label ``Ad10g,'' whose dust amplitude is drawn instead from a power spectrum estimated from the \texttt{d10} dust simulation at 353~GHz, whose shape is shown in Figure~\ref{fig:dust_sync_chi2}. We construct this by interpolating the binned noiseless dust $B$-mode power spectrum at 353\,GHz in our sky region.
    
    \subsubsection*{Homogeneous SED}
    \label{sssec:hom-gauss}
    This scenario corresponds to the \textit{Simple Gaussian} row in Table~\ref{tab:all_fg_sim_summary}, 
    where spatially constant SED parameters are assumed across the sky. Specifically, the foreground
    maps at the pivot frequency are scaled to each SO channel using a single set of dust and synchrotron 
    SED spectral parameters uniform across the sky. 
    The case is equivalent to having $x_{\beta_d}^{\rm G} = x_{T_d}^{\rm G} = x_{\beta_s}^{\rm G} = 0$ in 
    Equation~\ref{eq:decorr_param_scaling}.
    We generate 500 Gaussian realizations for covariance
    matrix estimation and an additional 100 realizations for component separation testing in the simplest 
    Gaussian case.

    \subsubsection*{Inhomogeneous SED}
    \label{sssec:inhom-gauss}
    To compare with non-Gaussian foregrounds from the \texttt{PySM} model (Section~\ref{subsec:real-foreground}), we also generate simulations with anisotropic SED parameters.
    Specifically, the \textit{$\beta_d$ Gauss} scenario in Table~\ref{tab:all_fg_sim_summary} corresponds to Gaussian simulations with inhomogeneous $\beta$ that mimic the foreground SEDs of the ``$\beta_d$-Spatial'' scenario described in Section~\ref{subsec:s5d10-fg}.
    The $\beta_d$ and $\beta_s$ templates are drawn from a Gaussian field with a ``broken power-law'' angular power spectrum: 
    \begin{align}
        D_\ell^{\beta\beta, c} = \begin{cases}
            B_c\cdot\left(\frac{\ell}{80}\right)^{\gamma_c}&\ell \geq 15\\
            D_{\ell = 15}^{\beta\beta,c}&\ell < 15
        \end{cases}
    \end{align}
    where $c \in {d, s}$ refers to dust and synchrotron, respectively. The break at $\ell=15$ suppresses 
    large-scale power to reduce variance in the mean spectral indices over the SO mask region.
    The break also preserves the power-law form of the $\beta$ templates within the analysis 
    scale range $30 < \ell < 300$.
    We use the \texttt{synfast} routine in the \texttt{healpy} library to draw $\beta_d$ and $\beta_s$ templates as spin-0 fields, shifted by a mean value $\bar{\beta}_c$. The templates are based on power spectrum fits to the \textsf{PySM} \texttt{d10} (for dust) and \texttt{s5} (for synchrotron) models, with
    \begin{align*}
        B_d = 3.682\times 10^{-6}&&\gamma_d &= -2.455&\bar{\beta}_d &= 1.6\\
        B_s = 0.611\times 10^{-6}&&\gamma_s &= -2.970&\bar{\beta}_s &= -3.0.
    \end{align*}
    For dust, we fix the dust temperature to $T_d = 19.6$~K across the sky. The $\beta_d$ template is then rescaled according to Equation~\ref{eq:decorr_param_scaling} for $x_{\beta_d}^G\in[0, 12.5]$ in increments of 2.5.
    
\begin{figure}[t]
        \centering
        \includegraphics[width=\linewidth]{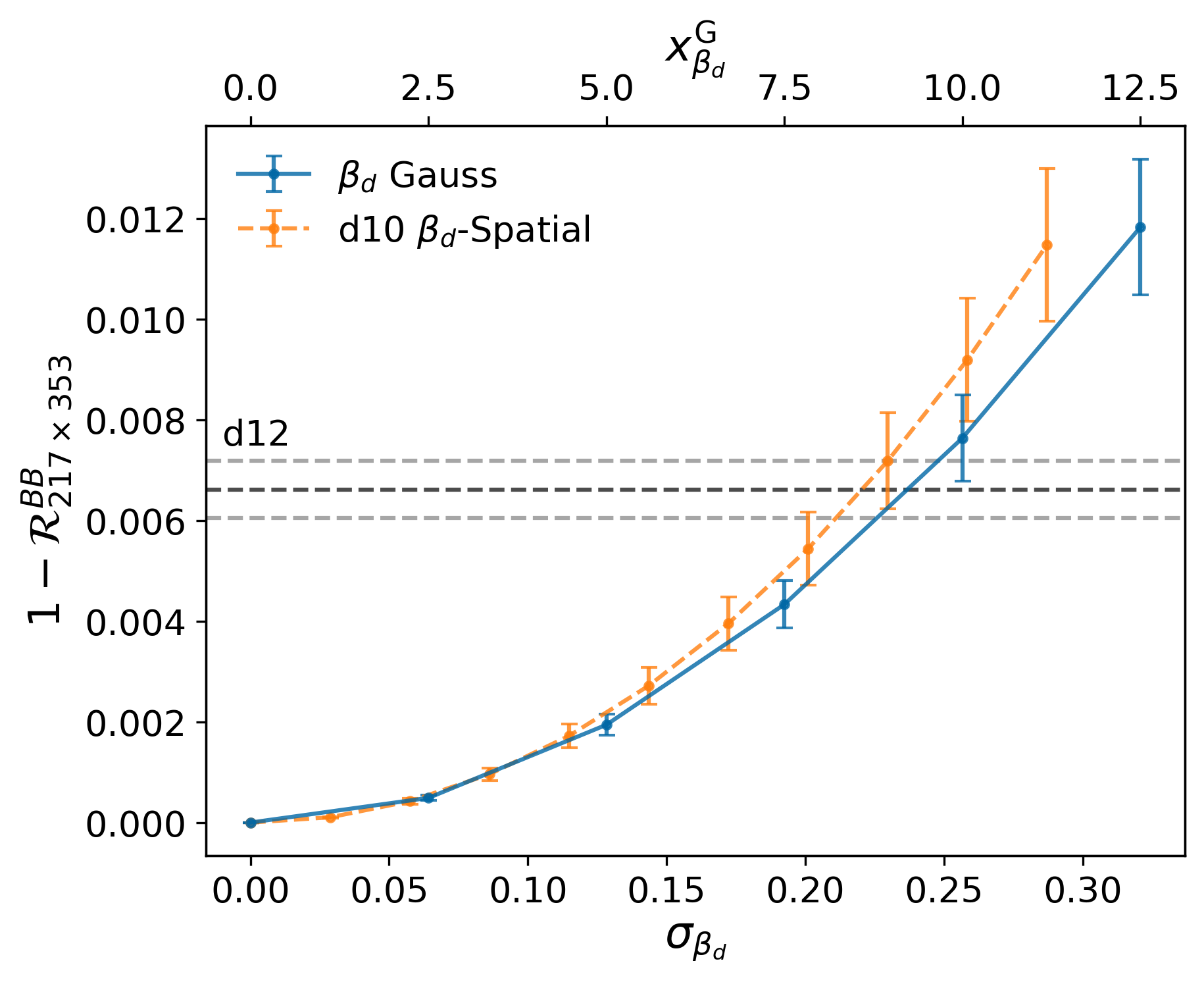}
        \caption{Frequency decorrelation over the Planck LR71 mask in the Gaussian simulations as a function of 
        dispersion in $\beta_d$ ($\sigma_{\beta_d}$, blue). The top x-axis label indicates the corresponding 
        scaling factor, $x_{\beta_d}^{\rm G}$. 
        This is compared to the \textit{$\beta_d$-Spatial} scenario (orange) from Figure \ref{fig:beta_d_scaling}. 
        As in Figure \ref{fig:beta_d_scaling}, the \texttt{d12} level is indicated.}
        \label{fig:gauss_decorr}
    \end{figure}

    The corresponding frequency decorrelation against $\beta_d$ scalings are shown in Figure~\ref{fig:gauss_decorr}, where the highest-dispersion Gaussian case has similar decorrelation levels to the \texttt{d10} model in the $\beta_d$-spatial configuration. The 
    non-Gaussianity of the \texttt{d10} $\beta_d$ template leads to additional frequency decorrelation compared
    to a Gaussian realization of $\beta_d$ with the same level of dispersion.
    The dust maps at the pivot frequency are scaled to the SO observation channels using the fixed dust temperature and the corresponding $\beta_d$ maps. For each value of $\beta$ dispersion, we generate 100 independent realizations. 

    \begin{figure}[t]
        \centering
        \includegraphics[width=\linewidth]{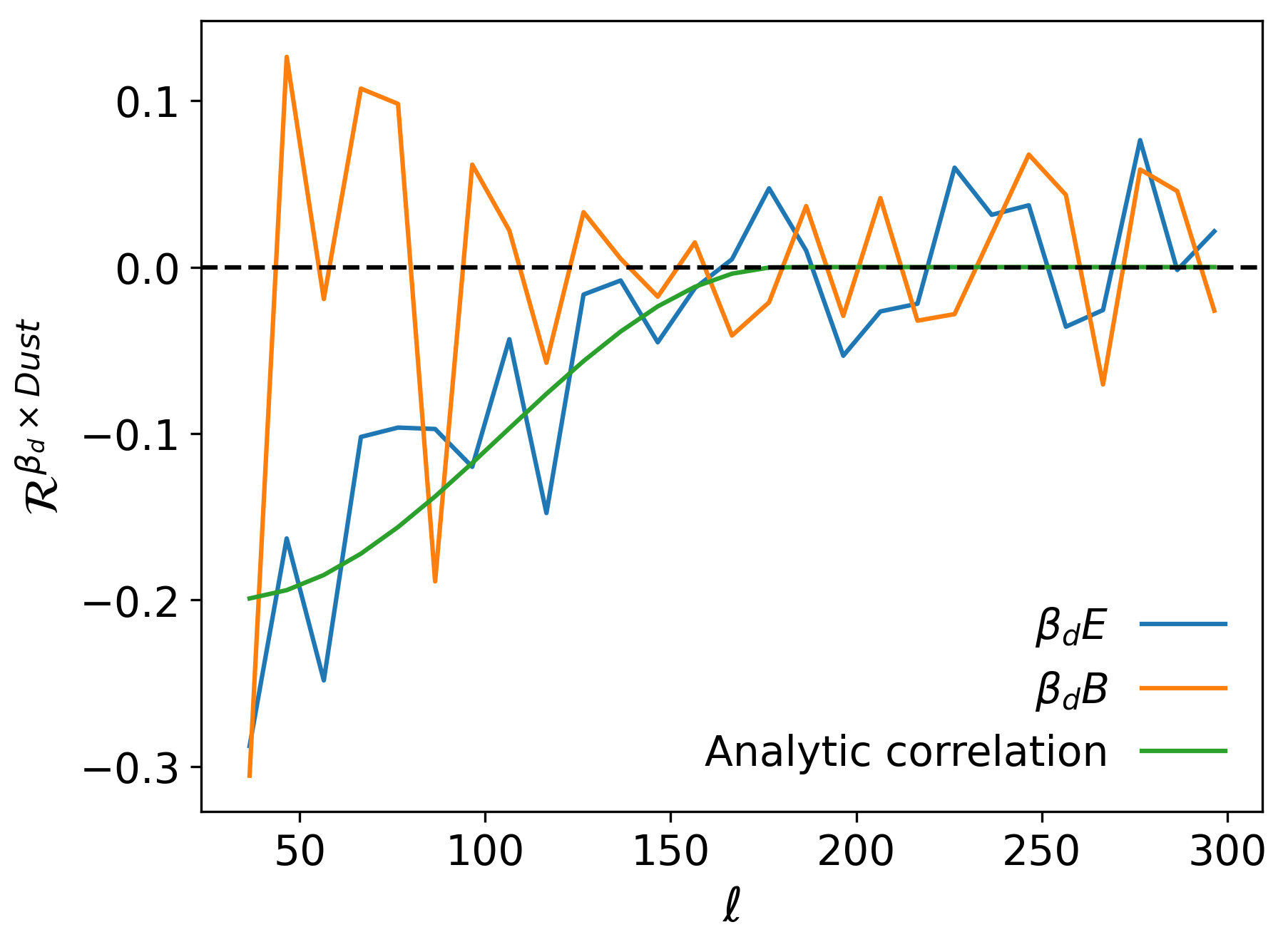}
        \caption{Comparison between the analytic correlation function $\rho_\ell$ (green, see Equation~\ref{eq:analytic-corr}) and the empirical correlation, $\mathcal{R}_\ell$, computed from the \texttt{d10} $\beta_d$ template map and the polarized \texttt{d10} $(Q,U)$ maps at 353~GHz, for $E$-mode (blue) and $B$-mode (orange) polarization. The analytic form is designed to approximate the scale-dependent behavior observed in the \texttt{d10} model.}
        \label{fig:analytic_correlation}
    \end{figure}
    
    Synchrotron $\beta$ maps are held fixed across all dust $\beta$ dispersion cases with 
    $x_{\beta_s}^{\rm G} = 1$. We generate 100 synchrotron realizations and reuse them for each of the Gaussian dust scenarios described above.

    In addition to component separation tests, these Gaussian simulations are used to construct covariance matrices described in Section~\ref{sec:cross_cl-covar}. These help quantify the impact of 
    neglecting frequency decorrelation in the covariance matrices.\\

    \begin{figure*}[t]
        \centering
        \includegraphics[width=\linewidth]{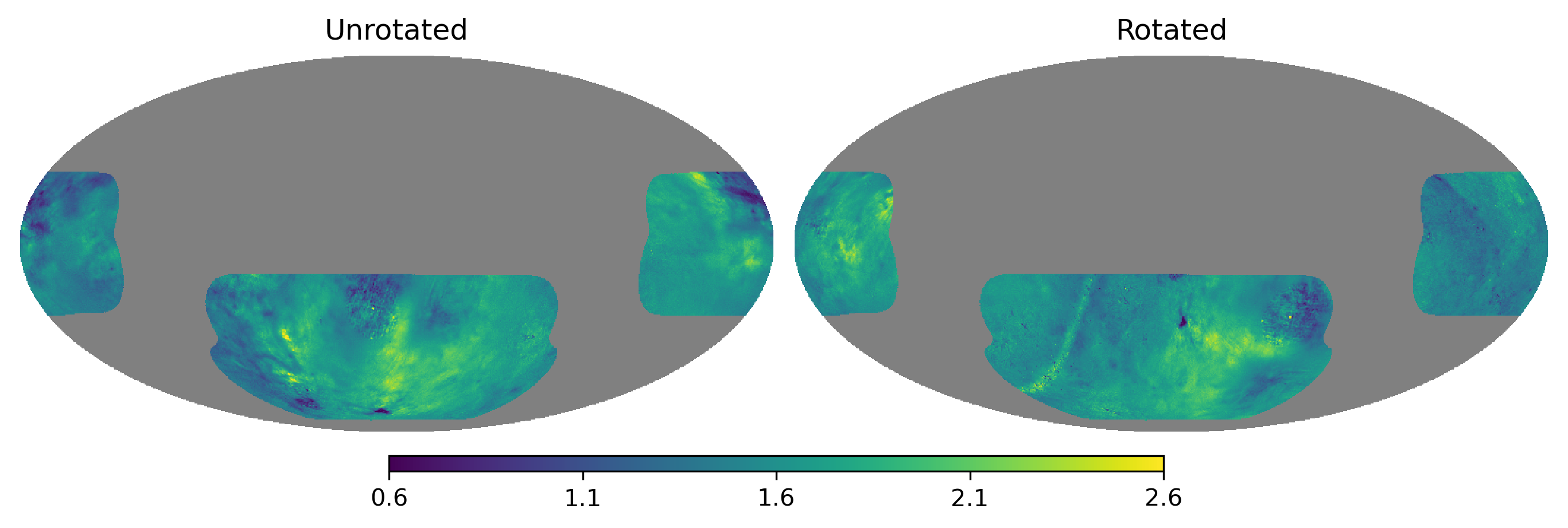}
        \caption{The $\beta_d$ template for the nominal unrotated \texttt{d10} simulation (left) compared with one of the rotated versions (right), within the SO mask. Both rotations exhibit pronounced non-Gaussian spatial structures, with the rotation shifting the location and 
        orientation of these features across the sky.}
        \label{fig:beta_rot}
    \end{figure*}

    To investigate the potential impact of correlations between $\beta_d$ and the dust amplitude map on estimates of the tensor-to-scalar ratio $r$, we also generate simulations that explicitly introduce such correlations. This corresponds to the \textit{$\beta$-Dust Corr.} scenario in Table~\ref{tab:all_fg_sim_summary}.
    Specifically, we jointly simulate the $\beta_d$ and dust amplitude using the \texttt{synfast} routine from \textsf{healpy}, with the following set of input power spectra:
    \begin{align*}
        C_\ell^{TT} &
        = C_\ell^{\beta\beta, d} & C_\ell^{TE} &= \rho_\ell \sqrt{C_\ell^{\beta\beta, d}C_\ell^{EE, d}}\\
        C_\ell^{EE} &= C_\ell^{EE, d} & C_\ell^{EB} &= 0\\
        C_\ell^{BB} &= C_\ell^{BB, d} & C_\ell^{TB} &= \rho_\ell \sqrt{C_\ell^{\beta\beta, d}C_\ell^{BB, d}}.
    \end{align*}
    This produces a spin-0 $\beta_d$ map that is correlated with the $Q$ and $U$ dust amplitude maps. Here $\rho_\ell$ is a scale-dependent correlation function:
    \begin{align}
        \label{eq:analytic-corr}
        \rho_\ell = \begin{cases}
            -0.2 & \ell < 30\\
            -0.1\cdot \left(\cos(\frac{\ell - 30}{180 - 30}\pi) + 1\right) & 30\leq \ell \leq 180\\
            0 & \ell > 180.
        \end{cases}
    \end{align}
    This functional form of the correlation is motivated by the observed relationship between the dust $E$-mode
    polarization and $\beta_d$ in the \texttt{d10} model. 
    We generate 100 realizations of simulations that incorporate this artificial correlation, where we fix the $\beta_d$ dispersion at $x_{\beta_d}^{\rm G} = 12.5$. If the correlation affects $r$ bias, we
    expect its impact to be the strongest at maximum $\beta_d$ decorrelation. 
    
    Figure~\ref{fig:analytic_correlation} shows the analytic correlation function $\rho_\ell$ (green), alongside the empirical correlation $\mathcal{R}_\ell$ computed between the \texttt{d10} $\beta_d$ template and the polarized \texttt{d10} $(Q,U)$ maps at 353~GHz. The resulting $\mathcal{R}_\ell$ curves are shown separately for $E$-mode (blue) and $B$-mode (orange) polarization.
    Although the correlation between dust $B$-mode and $\beta_d$ is consistent with zero, as shown in 
    Figure~\ref{eq:analytic-corr}, we construct this 
    scenario with an imposed correlation to explicitly test the impact of a zeroth-order $\beta_dB$ correlation on the estimation of $r$.

    \subsubsection{\texttt{d10}-Gaussian mixture model}
    \label{subsec:sim-SBd10}

    We generate an additional suite of simulations to investigate what properties of the dust can induce bias 
    in the tensor-to-scalar ratio. 
    We construct hybrid simulations combining Gaussian and non-Gaussian templates for the dust 
    amplitude and $\beta_d$, in order to better identify which component's statistical non-Gaussianity is 
    primarily responsible for any bias in $r$.
    To isolate the dust contribution, we fix the synchrotron model to follow Gaussian simulations generated from power-law spectra and including spatially varying $\beta_s$, as described in Section~\ref{subsec:gauss_fg}. This setup captures realistic inhomogeneous SED effects. 

    We generate dust simulations where the $\beta_d$ and amplitude maps are independently taken from either the \texttt{d10} template (Section~\ref{subsec:real-foreground}) or the Gaussian dust simulations (Section~\ref{subsec:gauss_fg}). This allows for fully Gaussian, fully \texttt{d10}, or hybrid configurations combining elements from both. 

    To ensure consistent frequency decorrelation across cases, we fix the \texttt{d10} $\beta_d$ dispersion 
    scaling to $x_{\beta_d} = 1.6$ and scale the Gaussian $\beta_d$ accordingly to match this \texttt{d10} 
    frequency decorrelation between 353~GHz and 217~GHz. The dust temperature is fixed at $T_d = 19.6$~K for 
    all simulations.

    This simulation suite includes four combinations of dust amplitude and $\beta_d$ maps, summarized as the \textit{Ad10*Bd*} scenarios in Table~\ref{tab:all_fg_sim_summary}: 
    \begin{itemize}
        \item \textit{Ad10Bd10}: both amplitude and $\beta_d$ maps from the \texttt{d10} template,
        \item \textit{Ad10Bdg}: amplitude from \texttt{d10}, $\beta_d$ from a Gaussian simulation,
        \item \textit{Ad10gBd10}: amplitude from a Gaussian realization drawn from the \texttt{d10} template spectrum, $\beta_d$ from \texttt{d10},
        \item \textit{Ad10gBdg}: amplitude from a Gaussian realization drawn from the \texttt{d10} template spectrum, $\beta_d$ from a Gaussian simulation.
    \end{itemize}
     
    We generate 100 signal realizations for each configuration. For the configuration using the original \texttt{d10} amplitude and $\beta_d$ maps (\textit{Ad10Bd10}), only one dust realization is used. In this case, the CMB and synchrotron components vary across realizations, while the dust remains fixed.
    
    We also construct a variant of the \textit{Ad10Bd10} scenario, referred to as 
    \textit{Ad10Bd10 rots} in Table~\ref{tab:all_fg_sim_summary}, in which four different 
    coordinate rotations are applied to the $\beta_d$ template prior to constructing 
    frequency-dependent dust maps. These rotations intentionally misalign 
    the $\beta_d$ and dust amplitude templates, breaking their pixel-by-pixel correspondence 
    when computing SED scalings.
    This simulation suite leverages the inhomogeneous structure of the $\beta_d$ template and 
    is designed to probe the impact of non-Gaussianity in $\beta_d$, and higher-order correlations 
    between the dust amplitude and $\beta_d$ templates, on any resulting bias in $r$. The $\beta_d$ template for the nominal unrotated \texttt{d10} simulation is shown in Figure \ref{fig:beta_rot} together with one of the rotated versions.

%% file: pipeline.tex
    \section{Pipeline}
    \label{sec:cross_cl}
    A widely used approach for foreground modeling in CMB $B$-mode analyses is the cross-spectrum ($C_\ell$-based) likelihood formalism, which models CMB and foreground components parametrically in harmonic space. This method constructs a data vector from all auto- and cross-power spectra between frequency channels, $C_\ell^{\nu\nu'}$, and uses it in conjunction with an estimated covariance matrix to perform joint inference of cosmological and foreground parameters. The covariance captures contributions from sample variance, instrumental noise, and foreground uncertainties.

    This approach is particularly well suited to ground-based experiments, where time-domain filtering and non-uniform, non-white noise introduce non-trivial pixel-pixel correlations that can be more easily treated in harmonic space. The cross-$C_\ell$ method has been successfully applied to real data analyses, including the tightest constraint on the tensor-to-scalar ratio, $r < 0.032 - 0.036$, from BICEP/Keck data combined with Planck and WMAP data \citep{bicepkeck_2021_r_constraint, Tristram_2022_bicep_npipe_r}. It has also been benchmarked against alternative component separation approaches in the context of forecasting and simulation-based validation for SO \citep{Wolz_2023_pipeline_comparison}.

    The version of the pipeline used in this work is implemented in the open-source \textsf{BBPower} software\ftnote{\url{https://github.com/simonsobs/BBPower}}, and is described in detail in \citet{Abitbol_2021_SO_calib_requirement, azzoni_2021_bbpower_moment, azzoni_2023_hybrid_bbpower} and  \citet{Wolz_2023_pipeline_comparison}. This pipeline consists of three major stages: power spectra computation, 
    covariance matrix estimation, and parameter inference. 

    \subsection{Power Spectra Computation}

    In this stage, we construct the data vector by computing auto- and cross-frequency $B$-mode power spectra of the six frequency channels from the simulated sky maps using the \textsf{NaMaster} package \citep{Alonso_2019_namaster}. 
    To mitigate $E$-to-$B$ leakage caused by masking, we apply the $B$-mode purification feature in 
    \textsf{NaMaster} \citep{Lewis_2001_masking_EB_leakage}. We adopt the analysis mask used in
    \citet{Wolz_2023_pipeline_comparison}, which is based on the hit-count forecast for SO from 
    \citet{SO_2019_forecast} smoothed with a 10-degree ``C1" apodization. Our analysis 
    targets the multipole range $30 \leq \ell \leq 300$ 
    of the SO SATs, which captures the recombination bump in the primordial $B$-mode signal, peaking at 
    $\ell\simeq 80$. 
    We assume that, in real data, the largest angular scales are significantly affected by atmospheric noise and filtering, while the smaller scales, primarily targeted by the SO-LAT and relevant for lensing $B$-modes, are not expected to carry significant primordial $B$-mode signal.
    We bin the power spectra uniformly with $\Delta\ell = 10$. 

    To mitigate instrumental noise bias, we
    split each simulation realization into multiple maps sharing the same sky signal (CMB and foregrounds) but with independent noise realizations (see Section~\ref{sec:sim-noise}).
    The polarized cross-power spectra are estimated by averaging over all unique cross-splits, excluding auto-split spectra to avoid noise bias. Specifically, for each pair of frequency channels ($\nu,\nu'$), we compute:
    \begin{align}
        \label{eq:x_split_mean}
        C_\ell^{\nu\nu'} \equiv \frac{1}{S(S - 1)}\sum_{\{s, s'| s < s'\}}C_\ell^{\nu s, \nu's'}~,
    \end{align}
    where $C_\ell^{\nu\nu'}$ is the cross spectrum between channel $\nu$ and $\nu'$, $S$ is the number of splits, and
    $C_\ell^{\nu s, \nu's'}$ is the cross channel power spectrum calculated using channel $\nu$ from split $s$ and 
    channel $\nu'$ from split $s'$. In this work, we use four independent splits per simulation, resulting in six cross-spectra per channel pair.  For $N = 6$ SO SAT frequency channels, the full data vector for one simulation contains 
    $N(N + 1) / 2 = 21$ spectra, or 567 data points, combining all cross- and auto-spectra.
    
    \subsection{Covariance Estimation}
    \label{sec:cross_cl-covar}
    
To estimate the bandpower covariance matrix, we consider three approaches, investigating the impact of including frequency decorrelation and deviations from a pure power-law shape in the foreground power spectra.
\begin{itemize}
\item In the first approach (`Cov1'), we use 500 power-law Gaussian simulations described in Section~\ref{sec:sims} and labeled `Simple Gauss' in Table \ref{tab:all_fg_sim_summary}. 
Each block of the covariance matrix corresponds to a pair of cross-frequency spectra. Consistent with findings in \citet{azzoni_2021_bbpower_moment}, the covariance is dominated by diagonal elements within each block.
    To suppress statistical noise due to the finite number of simulations, we retain only the main diagonal and six nearest off-diagonal elements within each block, setting all other entries to zero. This effectively preserves correlations between each bin and its three closest neighbors on each side. Given the width of our bandpower bins, this corresponds to retaining correlations within $\Delta\ell = 30$. Assuming that correlations beyond this range are negligible, the impact of this truncation is expected to be minimal. 

    This simple covariance matrix assumes no frequency decorrelation in each foreground component and power-law foreground power spectra. 
    These assumptions holds reasonably well for simpler foreground simulations used in previous studies 
    \citep[e.g.,][]{Wolz_2023_pipeline_comparison}, but break down under more complex foregrounds such as the \texttt{d10} dust
    model. When frequency decorrelation and the dust spectrum shape are not accounted for in the covariance matrix, 
    the resulting fit yields inaccurate $\chi^2$ values.
  
    \item A set of `Cov2' matrices include the effects of frequency decorrelation. Specifically, `Cov2' matrices use \textit{$\beta_d$ Gauss} simulations whose synchrotron and dust $\beta$ templates are Gaussian realizations drawn from broken power laws.
    Each `Cov2' matrix is computed using 100 Gaussian realizations. They include only the zeroth-order (main diagonal) elements within each frequency block, to reduce statistical noise that becomes significant given the smaller number of simulations. The variant of Gaussian realizations used for this covariance are labeled `$\beta_d$ Gauss' in Table~\ref{tab:all_fg_sim_summary}, and these realizations have varying degrees of decorrelation. 

    \item A final `Cov3' matrix uses the Gaussian realizations labeled `Ad10gBdg' and includes both decorrelation at a single scaling ($x=1.6$) and the non-power-law spectrum template to better capture the signal. As with `Cov 2,' it is computed using 100 Gaussian realizations and includes only the main diagonal elements within each frequency block.
    \end{itemize}

    \subsection{Parameter Inference}
    \label{sec:param_inf}

    \begin{table*}[t]
        \centering
        \hspace{-100pt}
        \begin{tabular}{c||ccccccccccccc}
            \hline\hline
            Parameters& $\alens$& $r$& $A_d$& $A_s$& $\alpha_d$& $\alpha_s$& $\epsilon_{ds}$& $\beta_d$& $\beta_s$& 
            $B_d$ & $B_s$ & $\gamma_d$ & $\gamma_s$\\
            \hline
            Type & TH & TH & TH & TH & TH & TH & TH & G & G & TH & TH & TH & TH\\
             Mid & - & - & - & - & - & - & - & 1.6 & -3.0 & 5 & 0 & -3  & -3\\
            Width & - & - & - & - & - & - & - & 0.1 & 0.3 & 10 & 5& 0.5 & 0.5\\
            \hline\hline
        \end{tabular}
        \caption{Priors adopted throughout our analysis for the free parameters. The type of prior is either tophat (TH) or Gaussian (G). ``Mid'' indicates the central point of 
        each prior distribution. ``Width'' means the half-width for top-hat priors and standard deviation for Gaussian priors. The parameters are 
        defined in Section \ref{sec:param_inf}. Empty Mid and Width entries indicate that the tophat prior is sufficiently broad to have no impact on the MCMC chains.
        }
        \label{tab:param-prior}
    \end{table*}
    
    We approximate the cross-spectrum likelihood as Gaussian and estimate the posterior distribution using Markov Chain Monte Carlo (MCMC) sampling. While the exact power spectrum likelihood is inherently non-Gaussian \citep[e.g.,][]{Hamimeche_2008_HL_likelihood}, the Gaussian approximation is well justified---as expected from the central limit theorem---in the limit of large sky coverage and a high number of independent modes per bandpower bin, 
    particularly over the analysis scale range $30 < \ell < 300$.
    Given the observing strategy and sky coverage of the SO SATs, and following previous analyses that adopt this assumption \citep{Planck_2014_ps_likelihood, Planck_2018_cosmo_params, azzoni_2021_bbpower_moment, azzoni_2023_hybrid_bbpower, Wolz_2023_pipeline_comparison}, we consider the Gaussian likelihood approximation to be sufficient for this work.

    Our parametric signal model includes contributions from the CMB and diffuse Galactic foregrounds, with $D_\ell = D_\ell^{\CMB} + D_\ell^{\fg}$. 
    The $B$-mode power spectrum is modeled using two parameters: the tensor-to-scalar ratio $r$ and a rescaling parameter for the lensing amplitude $A_{\textrm{lens}}$,
    \begin{align}
        \label{eq:cmb_model}
        D_\ell^{\CMB} = \alens D_\ell^{\textrm{lens}} + rD_\ell^r,
    \end{align}
    where $D_\ell^{\textrm{lens}}$ is the lensing-induces $B$-mode power spectrum, and $D_\ell^r$ is the primordial 
    $B$-mode power spectrum with $r = 1$. Both templates are computed using the Python package \textsf{CAMB} \citep{Lewis_2002_camb} with the best-fit $\Lambda$CDM cosmological parameters from \citet{Planck_2018_cosmo_params}.

    The foreground contribution to the power spectrum consists of synchrotron, dust, and their cross-correlation,
    \begin{equation}
        D_\ell^{\fg,\nu_i\nu_j} = D_\ell^{s,\nu_i\nu_j} + D_\ell^{d,\nu_i\nu_j} + D_\ell^{ds, \nu_i\nu_j}.
    \end{equation}
    At a given frequency, $D_\ell^s$ and $D_\ell^d$ are modeled as power law angular power spectra defined 
    in Equation~(\ref{eq:fg_ps}).
    Here we recall from Figure~\ref{fig:dust_sync_chi2} that this model is not a good fit 
    to the power spectrum of the \texttt{s5} synchrotron map at 27~GHz at the largest angular scales.

    To model the frequency dependence ($\nu_i\nu_j$), we adopt the following scaling:
    \begin{align}
        \label{eq:fg_freq_depend}
        D_\ell^{c,\nu_i\nu_j} = S_{\nu_0^c}^{c, \nu_i}S_{\nu_0^c}^{c, \nu_j}D_\ell^{c,\nu_0\nu_0}.
    \end{align}
    Here $S_{\nu_0}^{c, \nu}$ is the SED of the component $c\in\{s, d\}$ defined in Equation~(\ref{eq:dust_sed})
    and (\ref{eq:sync_sed}), and the pivot frequencies for synchrotron and dust 
    are $\nu_0^s = 23$~GHz and $\nu_0^d = 353$~GHz, respectively. 
    The cross-correlation between dust and synchrotron is modeled as
    \begin{align}
        \label{eq:ds_corr}
        D_\ell^{ds} = \epsilon_{ds}\sqrt{D_\ell^sD_\ell^d} ,    
    \end{align}
    with $\epsilon_{ds}$ a free parameter controlling the correlation amplitude \citep{Choi_2015_ds_corr}. 

    The baseline model therefore includes nine free parameters:
    \begin{align*}
        \{\alens, r, A_d, A_s, \alpha_d, \alpha_s, \epsilon_{ds}, \beta_s, \beta_d\},
    \end{align*}
    resulting in 558 degrees of freedom when fitting the data ($N_{\rm data} - N_{\rm params} = 567-9$).
    
    This baseline pipeline assumes no spatial variation in foreground spectral parameters. To relax this assumption, we adopt the minimal moment-expansion approach from \citet{azzoni_2021_bbpower_moment}, which has been shown to yield the tightest unbiased constraints on $r$ in the more complex foreground simulations explored by \citet{Wolz_2023_pipeline_comparison}. This extension incorporates spatial 
    variation of $\beta_d$ and $\beta_s$ as two power laws,
    \begin{align}
        C_\ell^{\beta_c\beta_c} = B_c\left(\frac{\ell}{\ell_0}\right)^{\gamma_c}, \, \, \, \, c\in\{d, s\}
    \end{align}
    which introduces four extra free parameters:
    \begin{align*}
            \{B_d, B_s, \gamma_d, \gamma_s\}.
    \end{align*}
    Here the $B$'s are the amplitudes of power laws at pivot scale $\ell_0 = 80$, and the $\gamma$'s are the indices\ftnote{See Equations (2.7), (2.16) and (2.17) in \citet{azzoni_2021_bbpower_moment} for 
    explicit expressions of the moment expansion terms.}.
    Since the power law index $\gamma_c$ is poorly constrained at small amplitude $B_c$, we apply an upper prior bound to prevent divergence in the MCMC sampling. This scenario has 554 degrees of freedom ($567-13$).

    We use the \textsf{emcee} package \citep{Foreman_mackey_2013_emcee} to perform MCMC sampling of the joint posterior distribution of the free parameters. The priors on all parameters are summarized in
    Table~\ref{tab:param-prior}. The analysis in \citet{bicepkeck_2021_r_constraint} showed that a prior on $\beta_d$ is no longer needed; we do include it here but find that the recovered posterior distributions for $\beta_d$ are much narrower than the prior width, indicating that the choice of prior does not significantly impact our results.

    To explore dust's deviations from a simple power-law angular spectrum (Equation~\eqref{eq:fg_ps}), we also test a model with the template shape used in the Ad10g simulation described in \ref{sec:sim-fg}, constructed from the binned noiseless dust $B$-mode power spectrum at 353\,GHz. We introduce an overall amplitude scaling parameter for the template, but do not allow its slope to vary. By estimating the template shape directly from the noiseless \texttt{d10} simulation at 353~GHz, we effectively reduce the number of degrees of freedom by at least 27 (the number of bandpower bins), since the total 280~GHz spectrum is signal-dominated and thus well described as a scaled version of the 353~GHz dust spectrum. 
    There are now 12 model parameters, so we approximate that there are $528$ degrees of freedom when fitting the template model to the \texttt{d10} simulated data ($567-27-12$). When this same template model is fit to the \texttt{Ad10g} simulated data, those data are drawn from the template spectrum and have scatter, so the degrees of freedom are not reduced in the same way; we approximate that there are $555$ degrees of freedom in this case ($567 - 12$).
   
    Of course, such noiseless templates cannot be constructed in practice; with real data we would expect to handle such deviations from power law behavior by constraining the dust amplitude in each bin separately. We also note that interpolation can suppress spectral features after rebinning, potentially introducing small biases.

%% file: result_section.tex
\section{Results}
\label{sec:result}

\subsection{Gaussian Scenario Results}
\label{subsec:result_gauss}

\begin{figure*}
    \centering
    \includegraphics[width=0.7\linewidth]{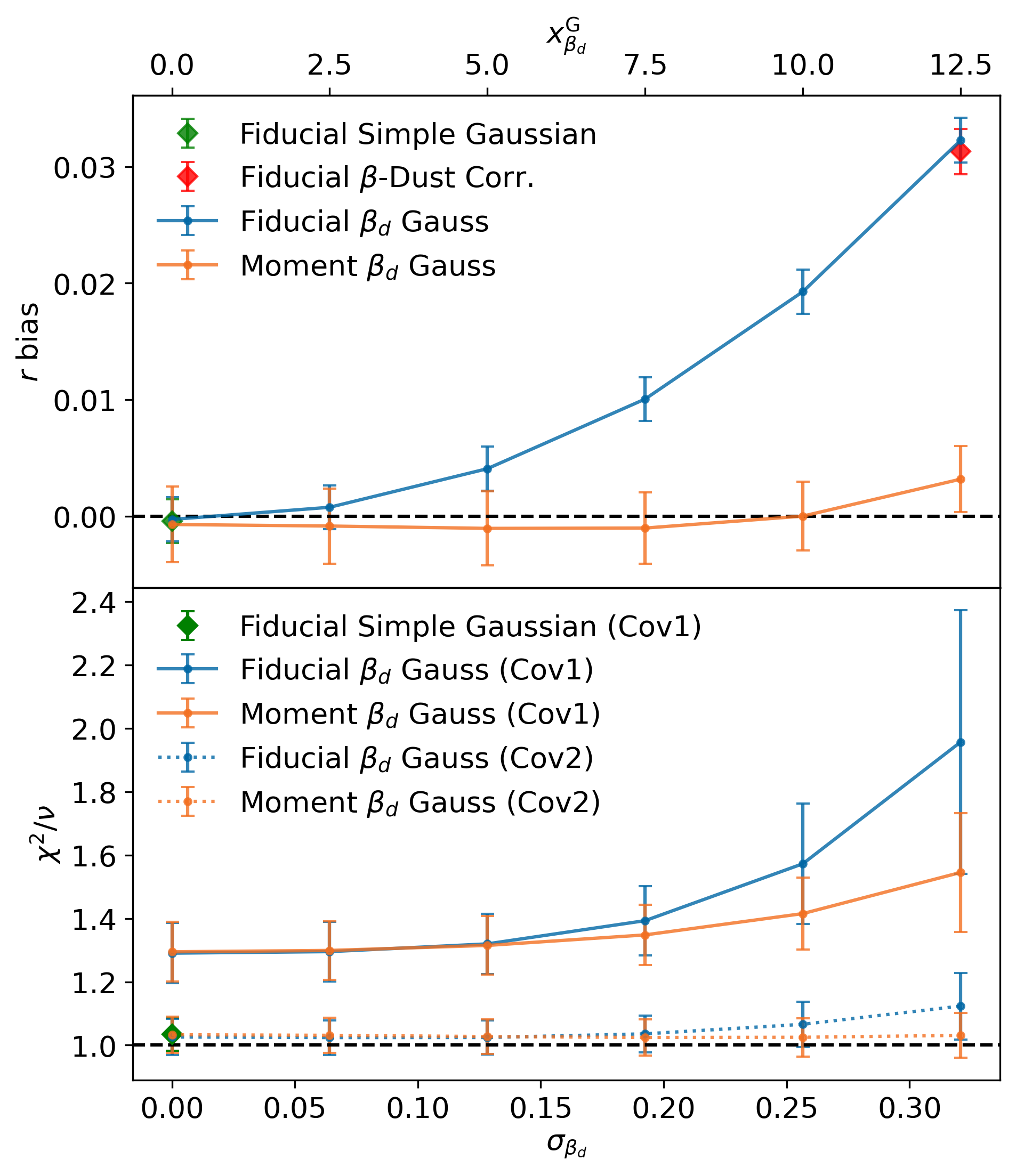}
    \caption{The bias in the recovered tensor-to-scalar ratio $r$ (top) and corresponding $\chi^2$ per degree of freedom (bottom) for Gaussian foreground simulations (\textit{$\beta_d$ Gauss}) with varying $\sigma_{\beta_d}$. The dashed black lines show the expected values for $r$ (top, $r_{\rm input} = 0$) and $\chi^2/\nu$ (bottom). The solid lines show results from the fiducial (blue) and moment (orange) pipelines using the fully correlated Gaussian covariance (Cov1). Using the moment model removes the substantial bias in $r$, and the dotted lines show that the $\chi^2/\nu$ is acceptable when using a covariance matrix that accounts for frequency decorrelation (Cov2). The green point shows the result for the \textit{Simple Gaussian} case with no spatial variation in indices; in this case $r$ is unbiased. The red point indicates the \textit{$\beta$-Dust Corr.} scenario, where correlations between the $\beta_d$ and the dust amplitude are included; this does not change the estimated $r$. Note that, even in the limit of $\sigma_{\beta_d}\rightarrow0$, the simulation exhibit a non-zero level of frequency decorrelation due to the spatial variability of the synchrotron spectral index, leading to the constant offset from $\chi^2/\nu=1$.}
    \label{fig:gauss_r_chi2}
\end{figure*}

We begin by validating the cross-spectrum pipeline using the \textit{Simple Gaussian} scenario with the corresponding Gaussian covariance matrix. As shown by the green point in Figure~\ref{fig:gauss_r_chi2} (top), the fiducial pipeline with no moment expansion recovers an unbiased estimate of the tensor-to-scalar ratio $r$, consistent with earlier results from \citet{azzoni_2021_bbpower_moment, azzoni_2023_hybrid_bbpower, Wolz_2023_pipeline_comparison}, when $x_{\beta_d}^G = x_{\beta_s}^G = 0$. The corresponding best-fit $\chi^2$ per degrees of freedom $\nu$ (bottom panel of the same figure) is close to unity, with PtE of 0.33, indicating good agreement between the model and the data.

Figure~\ref{fig:gauss_r_chi2} (top panel) illustrates that $r$ is increasingly over-estimated with increasing $\beta_d$ dispersion for the \textit{$\beta_d$ Gauss} scenario, where the $\beta_s$ dispersion is kept fixed at
$x_{\beta_s}^G = 1$ for a mild synchrotron frequency decorrelation. The solid lines indicate the estimated $r$ from the fiducial (blue) and the moment (orange) pipelines. In the fiducial case, the bias on $r$ increases with $\beta_d$ dispersion, up to a level of $\delta r=0.03$, while the $x_{\beta_d}^G = 0$ case remains unbiased, even despite the presence of spatially varying synchrotron $\beta_s$. The moment pipeline significantly reduces this bias across all dispersion levels, with results consistent with $r = 0$ within $1\sigma$ for all but the highest dispersion case. A small negative bias is observed for $x_{\beta_d}^G < 10$ at the $<0.25\sigma$ level, in agreement with the trends reported by \citet{azzoni_2021_bbpower_moment}, where it was attributed to prior and volume effects. As discussed there, this bias could be mitigated by adopting an appropriate prior (e.g., a Jeffreys prior) or by using a profile likelihood approach. 

The red point in the top panel of Figure~\ref{fig:gauss_r_chi2} corresponds to the \textit{$\beta$-Dust Corr.} scenario, in which $\beta_d$ is correlated with the dust amplitude (Section~\ref{subsec:gauss_fg}). Specifically, the correlation is imposed at the lowest (linear) order---we do not constrain higher-order correlation, 
such as those between $\beta_d^2$ and the dust amplitude, which could also influence the observed bias.
The bias on $r$ recovered with the fiducial cross-$C_\ell$ pipeline in this scenario is statistically consistent with the corresponding \textit{$\beta_d$ Gauss} case (blue), where no such correlation is present, even at high levels of frequency decorrelation.
This result suggests that correlations between $\beta_d$ and the dust amplitude have a negligible impact on the inference of $r$ under current assumptions.

The bottom panel of Figure~\ref{fig:gauss_r_chi2} shows the best-fit $\chi^2/\nu$ for the same set of simulations, estimated with the fiducial (blue) and moment (orange) cross-$C_\ell$ pipeline. When using the simple Gaussian covariance (solid lines), the $\chi^2$ is too high and increases with $\beta_d$ dispersion, indicating both a poor model fit due to unmodeled frequency decorrelation, and also an incorrect covariance matrix. Even the $x_{\beta_d}^G = 0$ case shows elevated $\chi^2$, driven by synchrotron decorrelation from spatially varying $\beta_s$. 

We find that incorporating frequency decorrelation into the covariance matrix, by including spatial variations of $\beta_d$ in the simulations used to compute the covariance, restores $\chi^2$ to values consistent with the number of degrees of freedom (dotted lines), underscoring the importance of including SED parameter dispersion in covariance modeling\ftnote{
    Although this method is not applicable to real data, the results highlight that modeling the impact of foreground frequency decorrelation in the covariance could become necessary if the level of decorrelation is significant. A detailed investigation of this effect is beyond the scope of this work and is left to future study.
}.
The $\chi^2$ values shown by the dotted curve are computed using the updated covariance matrix, while 
retaining the best-fit model obtained with the simple covariance. We demonstrate in 
Section~\ref{subsec:r_bias_source} that the updated covariance does not alter the best-fit model.
With the modified covariance, the moment pipeline maintains a stable $\chi^2$ across all cases, while the fiducial pipeline continues to show increased $\chi^2$ with growing $\beta_d$ dispersion. This is expected, as the fiducial model assumes a spatially uniform SED and is unable to account for $\beta_d$ variation.

\subsection{Realistic scenario results}
\label{subsec:result_real_r}

\begin{figure*}[th]
    \centering
    \includegraphics[width=\linewidth]{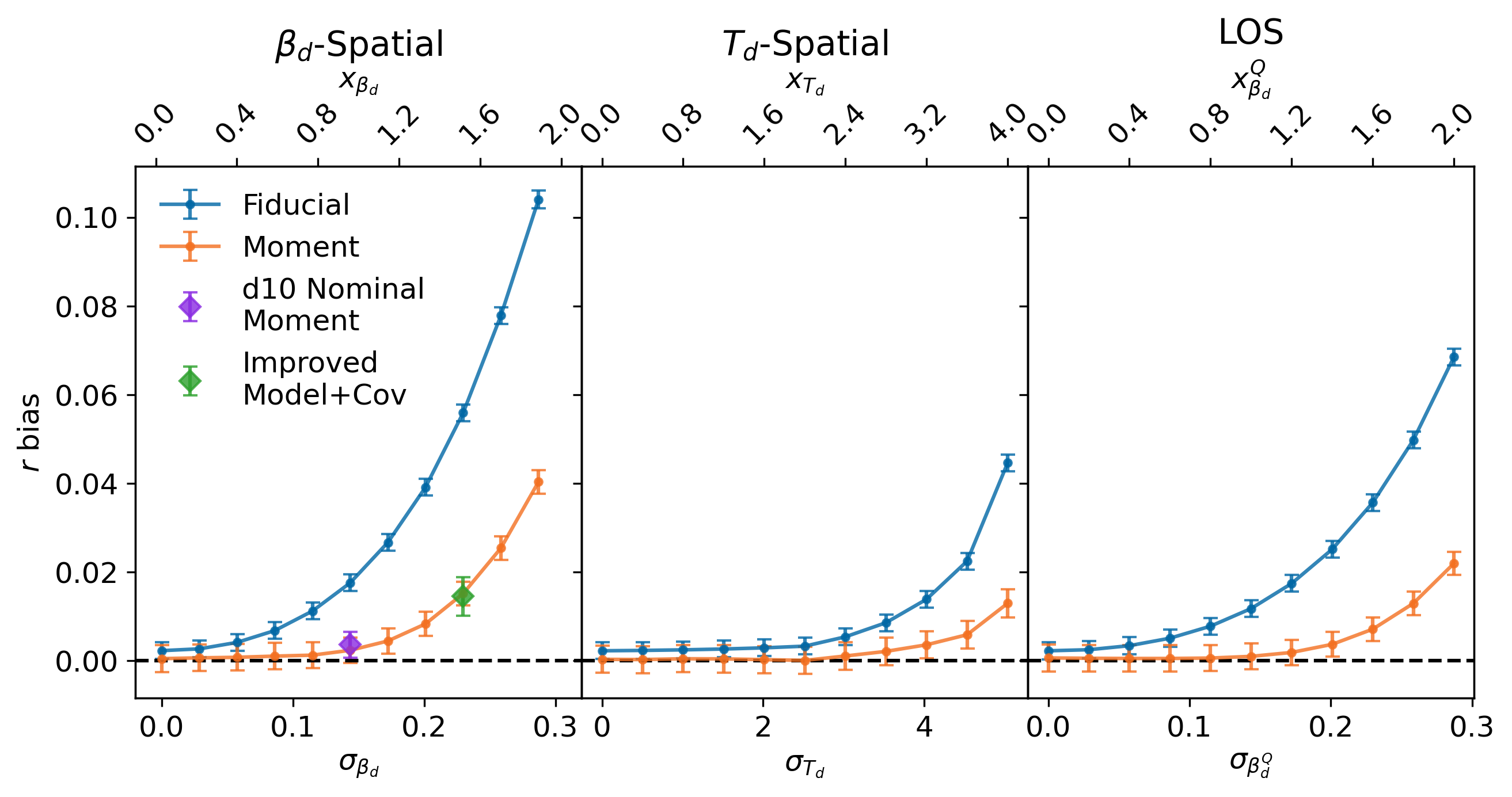}
    \caption{
    Recovered bias in $r$ as a function of SED parameter dispersion (x-axis) for the non-Gaussian simulations with three realistic decorrelation scenarios: \textit{$\beta_d$-spatial} (left), \textit{$T_d$-spatial} (center), 
    and \textit{LOS} (right). Results are shown for the fiducial (blue) and moment method (orange), using the \texttt{d10} dust and \texttt{s5} synchrotron models. The fiducial pipeline shows an increasing bias in $r$ as a function of dispersion, while the moment method mitigates bias at low to moderate levels---but unlike the Gaussian case shown in Figure \ref{fig:gauss_r_chi2}, it does not remove it at the highest dispersion levels. The left-hand \textit{$\beta_d$-spatial} case yields the largest bias for a given degree of decorrelation. For reference, the violet 
    point in the left panel notes the result for the \textit{\texttt{d10} Nominal} scenario using the moment pipeline. The estimated value for
    $r$ is consistent with the $x_{\beta_d}$ case of the \textit{$\beta_d$ Spatial} scenario when we 
    improve the modeling (angular spectral template) and the covariance 
    (Cov3, including frequency decorrelation and angular spectral template), as shown by the ``Improved Model+Covariance'' case (green, left panel) for the $x_{\beta_d}=1.6$ case, referred to in the text as \textit{Ad10Bd10}.
    }
    \label{fig:realistic_r_result}
\end{figure*}

We test three decorrelation scenarios in which frequency decorrelation is induced by spatial variations in either $\beta_d$, $T_d$, or along the line of sight (LOS), as described in Section~\ref{subsec:s5d10-fg}. The relationship between the scaling parameter $x$ and the resulting decorrelation ($1 - \mathcal{R_\ell}^{217\times353}$) for each case was shown in Figure~\ref{fig:beta_d_scaling}. 

Figure~\ref{fig:realistic_r_result} summarizes the $r$ constraints obtained for each scenario using realistic foreground models — \texttt{d10} for dust and \texttt{s5} for synchrotron (Section~\ref{subsec:real-foreground}) with the Cov1. In the limit of zero dust parameter dispersion (i.e., $x_{\beta_d} = x_{T_d} = 0$), the fiducial cross-spectrum pipeline yields a non-negligible bias on $r$ across all scenarios. This bias originates from synchrotron decorrelation due to spatial variations in $\beta_s$ present in the \texttt{s5} model. In contrast, the minimal moment-extension of \texttt{BBPower} recovers unbiased $r$, demonstrating that modeling SED variations with a second-order expansion in $\beta_s$ is sufficient in this regime.

In the \textit{$\beta_d$-spatial} and \textit{LOS} scenarios, we find that the bias on $r$ in the fiducial pipeline increases with $\beta_d$ or $\beta_d^Q$ dispersion, up to large values of order $r=0.1$. The moment-expansion pipeline substantially reduces this bias, remaining consistent with $r = 0$ at low dispersion levels. At higher dispersions, a substantial bias is still apparent, for $x_{\beta_d} \gtrsim 1.0$ in the \textit{$\beta_d$-spatial} case and $x_{\beta_d^Q} \gtrsim 1.2$ in the LOS case.

The \textit{$T_d$-spatial} scenario results in overall smaller biases in $r$ for both pipelines. For $x_{T_d} \lesssim 2.0$, the $r$ bias remains stable. At higher values, it begins to increase but remains below the bias observed at $x_{\beta_d} = 2.0$ in the \textit{$\beta_d$-spatial} case, despite both configurations inducing similar levels of frequency decorrelation (Figure~\ref{fig:beta_d_scaling}).
This comparison should be interpreted with caution due to the larger uncertainty in frequency decorrelation at $x_{T_d} = 4.0$ as shown in Figure~\ref{fig:beta_d_scaling}. A more reliable comparison can be made between $x_{T_d} = 3.2$ and $x_{\beta_d} = 1.0$, where the decorrelation levels are comparable and uncertainties are better constrained. In this regime, both pipelines yield consistent biases on $r$ across the scenarios, suggesting that the pipelines respond similarly to complexity induced by either $\beta_d$ or $T_d$ variations.\\

The parameter inference for these three scenarios uses the Cov1 covariance matrix, derived from Gaussian simulations with power-law spectra and no frequency decorrelation, which is known to be incorrect. As in the \textit{$\beta_d$ Gauss} scenario, this approach leads to a significant overestimation of the $\chi^2$. For zero dust decorrelation ($x_d = 0$), the $\chi^2$ per degree of freedom is 1.7, higher than the equivalent $\sim$1.3 in the \textit{$\beta_d$ Gauss} scenario. We find that this increase is primarily driven by the synchrotron component in the simulated data, which is based on the \texttt{s5} template. Unlike the Gaussian synchrotron model used in the \textit{$\beta_d$ Gauss} test, \texttt{s5} introduces stronger frequency decorrelation and its SED deviates from a pure power-law, worsening the mismatch with the model assumed to compute the simple covariance and thereby increasing the $\chi^2$. 

In this case, we find that incorporating synchrotron frequency decorrelation into the covariance\ftnote{`Cov2' that uses broken power law (BPL) $\beta_s$ template scaled by $x_{\beta_s}^{\rm G} = 6.4$ to match \texttt{s5} $\beta_s$ spatial dispersion, with a constant $\beta_d$.}
does not eliminate the $\chi^2$ excess when $x_{\beta_d}^{\texttt{d10}} = 0$, 
only reducing the $\chi^2$ per degree of freedom from 1.7 to 1.6. This behavior differs from that seen for the \textit{$\beta_d$ Gauss} scenario in Section~\ref{subsec:result_gauss}, suggesting that additional effects are contributing to the elevated $\chi^2$. As discussed in Section~\ref{subsec:real-foreground}, and shown in Figure~\ref{fig:dust_sync_chi2}, the angular power spectra of realistic dust and synchrotron (\texttt{d10} and \texttt{s5} here) depart from the pure power laws assumed in the Gaussian covariance. 

To assess the impact of incorrectly assuming a power-law model in the simulations used to compute the covariance, we test the Cov3 matrix that captures the shape of the angular power spectrum at the dust-dominant frequency, and accounts for the cross-frequency correlations. We do this for a specific case: the
 \textit{Ad10Bd10} simulation that combines power-law Gaussian synchrotron with \texttt{d10} dust, with $x_{\beta_d}^{\texttt{d10}} = 1.6$ (see Section~\ref{subsec:sim-SBd10}). 
Using the nominal power-law model and the moment-expansion, we find a significantly improved fit, with $\chi^2$ of $599 \pm 33$ for 554 degrees of freedom, with a PtE of 0.17\ftnote{Some residual $\chi^2$ mismatch may also arise from interpolation artifacts introduced by the binned template.}. 

We also test a case where we use the power spectrum template shape in the model as well as in the covariance matrix. Here we find $\chi^2$ of $564 \pm 33$ for 528 degrees of freedom and a PtE of 0.21
with the moment-expansion model. It appears that a power-law model for the dust, combined with a covariance matrix that accurately models the signal, is sufficient for the \texttt{d10} model at the adopted SO sensitivity. Higher sensitivity measurements may require more detailed modeling. Nonetheless, in some of our further investigations we choose to adopt the dust template shape in the model. 

We find that these adjustments in modeling method and covariance matrix have a limited effect on the recovered value for $r$,
as indicated by the green point in the first panel of Figure~\ref{fig:realistic_r_result}. The estimated $r$ 
using the modified Cov3 covariance and modeling method differs by less than $0.2\sigma$ compared to using Cov1 and 
power law dust modeling. However, we find that the uncertainty using Cov3 is larger than that for Cov1, due to the inclusion of the decorrelations. This effect is expected to be a function of decorrelation level, and will be explored in future work. We do not reproduce all of the results in Figure~\ref{fig:realistic_r_result} for the improved Cov3 covariance matrix.

\subsection{Diagnosing the Origin of Biases in $r$}
\label{subsec:r_bias_source}

\subsubsection{Role of $\beta_d$ Complexity}

We use a suite of mixture simulations that combine Gaussian and realistic models of 
dust and $\beta_d$ (introduced as \textit{Ad10*Bd*} scenarios in Section~\ref{subsec:sim-SBd10}) to investigate the source of the cosmological bias found, particularly when using the moment expansion method to model 
the simulated data.
This approach, detailed in \cite{azzoni_2021_bbpower_moment}, relies on several simplifying assumptions to characterize the multi-frequency power spectrum given spatial variations of foreground spectral parameters. Specifically, it assumes:
\begin{enumerate}
    \item no line-of-sight (LOS) decorrelation,
    \item spectral indices for each foreground component follow Gaussian distributions and are uncorrelated between different components,
    \item amplitudes and spectral indices of each foreground component are uncorrelated,
    \item amplitudes and spectral indices of each component are parametrized as power laws, 
    \item the expansion is truncated at leading order, 
\end{enumerate}
Based on the findings presented in the previous sections, certain assumptions, such as the correlation between amplitudes and spectral indices at lowest order (see Section~\ref{sssec:inhom-gauss} for discussion of the relevant simulations, and the red point in Figure~\ref{fig:gauss_r_chi2})
appear to be less critical sources of bias. 
Consequently, our analysis will primarily focus on assessing the impact of assumptions related to Gaussianity and the parametric form of the spectral models.

For testing, 
we adopt the Cov3 covariance matrix from the \textit{Ad10gBdg} simulation, which includes both frequency 
decorrelation and the correct angular power spectrum shape. The component separation step uses the dust 
power spectrum template as the dust power spectrum model.

\begin{figure*}[th]
    \centering
    \includegraphics[width=0.8\linewidth]{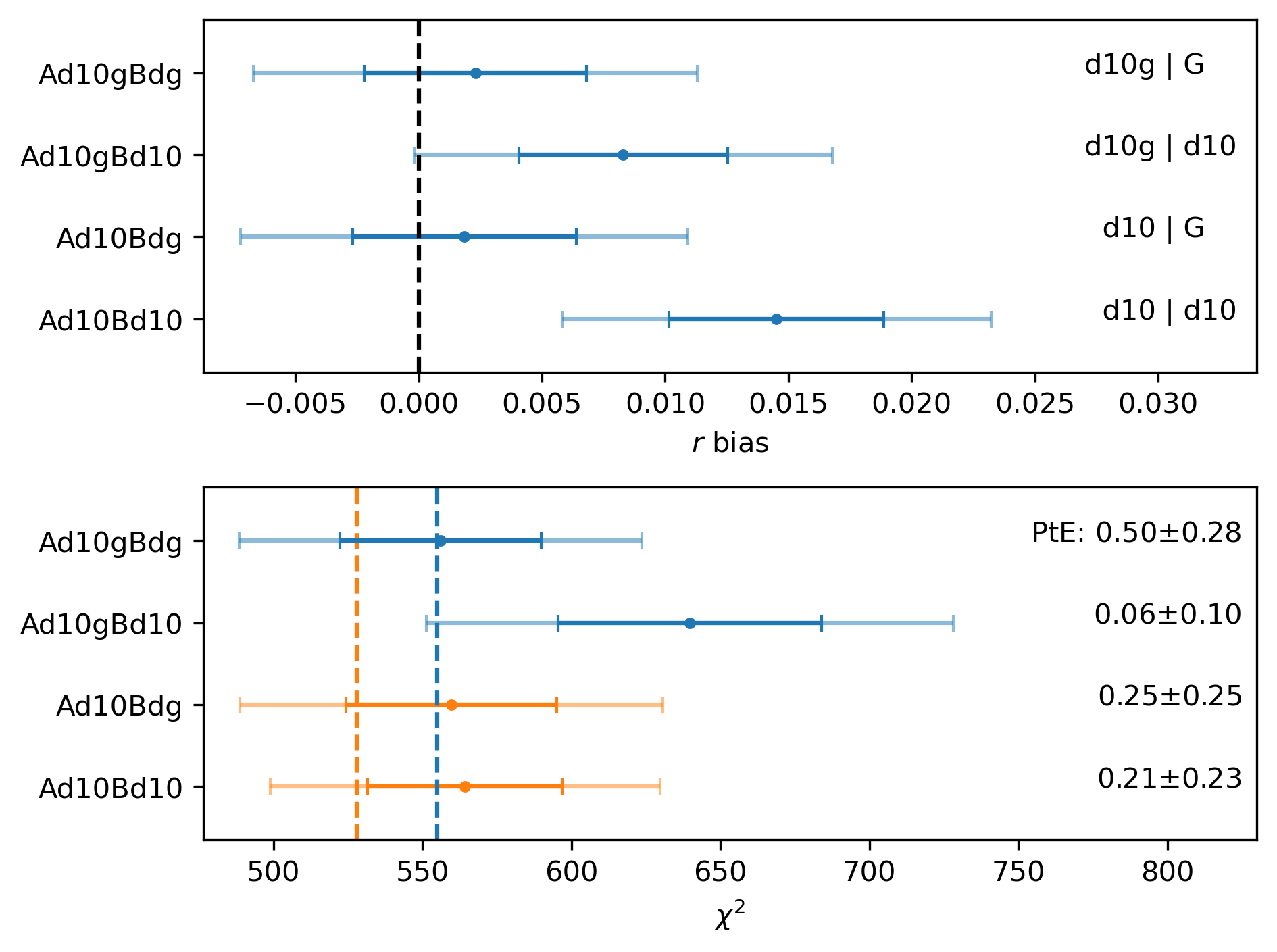}
    \caption{Top: $1\sigma$ and $2\sigma$ bounds for the estimated $r$ for four \texttt{d10} mixture model scenarios that combine Gaussian and non-Gaussian maps for the dust amplitude and emissivity. 
    The dust amplitude and $\beta_d$ templates are labeled next to each scenario (amplitude $|$ $\beta_d$), with the `g' indicating a Gaussian map. Biases in $r$ are only seen when the $\beta$ maps are non-Gaussian.
    Bottom: $2\sigma$ bounds for the $\chi^2$ values for the same scenarios. The degrees of freedom for \texttt{d10} (528, orange) and \texttt{d10g} (555, blue) amplitude templates are 
    distinguished by the coloring. The PtE values for each scenario are labeled on the right.}
    \label{fig:mixture-result}
\end{figure*}

Figure~\ref{fig:mixture-result} summarizes the $r$ bias ($\delta r$), $\chi^2$, and PtE values for each combination of dust amplitude and $\beta_d$ template. The top panel shows $2\sigma$ bounds of $r$ compared to the input $r=0$ (vertical dashed black line). 
Using the \textit{Ad10Bdg} and \textit{Ad10gBdg} scenarios,
we recover similar $r$ values of $0.0019\pm 0.0045$ and $0.0023\pm 0.0045$, respectively, which are statistically consistent with zero. This indicates that the bias on $r$ is not driven by non-Gaussianities in the realistic \texttt{d10} dust amplitude, consistent with the findings of \citet{Abril_Cabezas_2023_non_gauss_fg}.

The \textit{Ad10gBdg} scenario yields a $\chi^2$ of $556 \pm 34$ for 555 degrees of freedom, with a PtE of 0.50, as expected for a well-specified model. In comparison, the \textit{Ad10Bd10} scenario produces a slightly elevated $\chi^2$ of $564 \pm 33$ for 528 degrees of freedom. 
Both cases demonstrate that incorporating a spectral template into the modeling, alongside a simulation-based covariance that reflects the same foreground complexity, is effective in mitigating $\chi^2$ mis-estimation. Moving forward, while Gaussian realizations remain useful for covariance construction, future analyses should also consider analytical or semi-analytical covariance formulations to further suppress biases arising from mismatches in foreground modeling \citep[e.g.,][]{Knox_1995_covariance, Akins_2024_act_semi_analytical_covar, Abril_Cabezas_2023_non_gauss_fg}.

In contrast to the two Gaussian $\beta_d$ scenarios, we find that using the \texttt{d10} $\beta_d$ template introduces a bias in $r$ regardless of the dust amplitude model. Both the \textit{Ad10gBd10} scenario, which uses a Gaussian dust amplitude computed from the \texttt{d10} power spectral template, and the \textit{Ad10Bd10} scenario,
which uses the non-Gaussian Planck GNILC amplitude template, yield significant biases on $r$ of $0.008$ and $0.014$,
respectively. This indicates that the primary source of the bias is the \texttt{d10} $\beta_d$ template itself. As shown in Section~\ref{subsec:result_gauss}, this effect cannot be attributed to lowest-order correlations between dust intensity and $\beta_d$, suggesting that higher-order statistical correlations between dust amplitude and $\beta_d$ as well as non-Gaussian properties in the \texttt{d10} $\beta_d$ field may be responsible.

Both \texttt{d10} $\beta_d$ scenarios also exhibit worse $\chi^2$ values, likely driven by poorer agreement with the cross-frequency spectra. The GNILC-derived $\beta_d$ template appears less well-behaved, due to its inherent non-Gaussianity and higher-order correlation with dust amplitude, compared to Gaussian realizations drawn from a broken power-law model. These differences manifest in the cross-channel power spectra, degrading the quality of the fit and contributing to the increased biases in $r$.

\subsubsection{Bias mechanism}

In Section~\ref{subsec:result_gauss}, we showed that the first-order correlations between $\beta_d$ and dust amplitude do not contribute significantly to the bias in $r$. However, higher-order 
correlations, arising from either fitting degeneracies or physical processes, may still contribute.

\begin{figure*}[ht]
    \centering
    \includegraphics[width=0.8\linewidth]{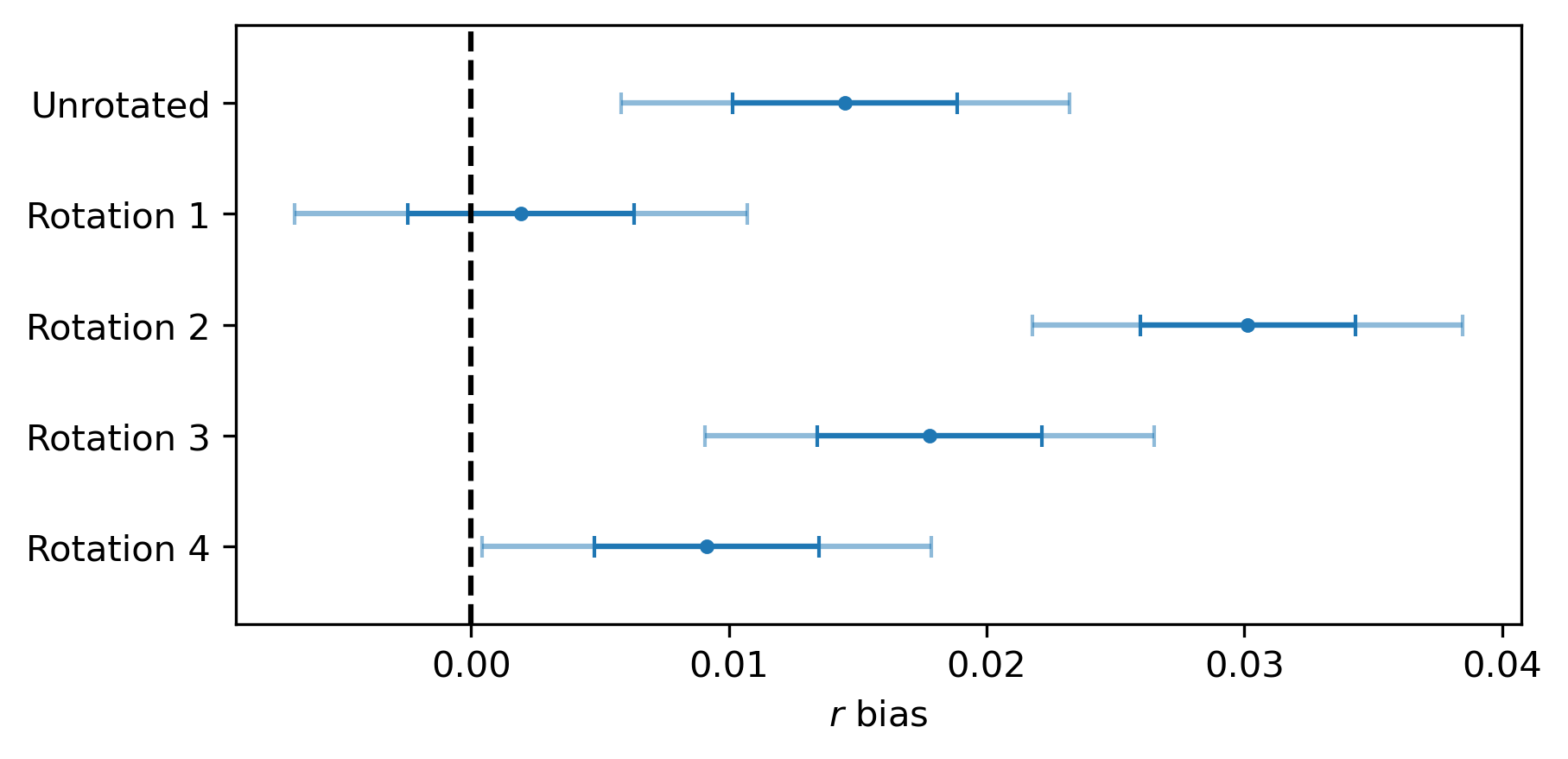}
    \caption{$1\sigma$ and $2\sigma$ bounds on the estimated $r$ found using the moments method for the different \textit{Ad10Bd10 rots} cases. 
    Four different coordinate rotations of the $\beta_d$ map are compared to the standard unrotated \texttt{d10} case; the choice of rotation of the non-Gaussian $\beta_d$ field has a large impact on the estimated $r$. 
    }
    \label{fig:d10_rot_r_bias}
\end{figure*}

\begin{figure*}[ht]
    \centering
    \includegraphics[width=0.75\linewidth]{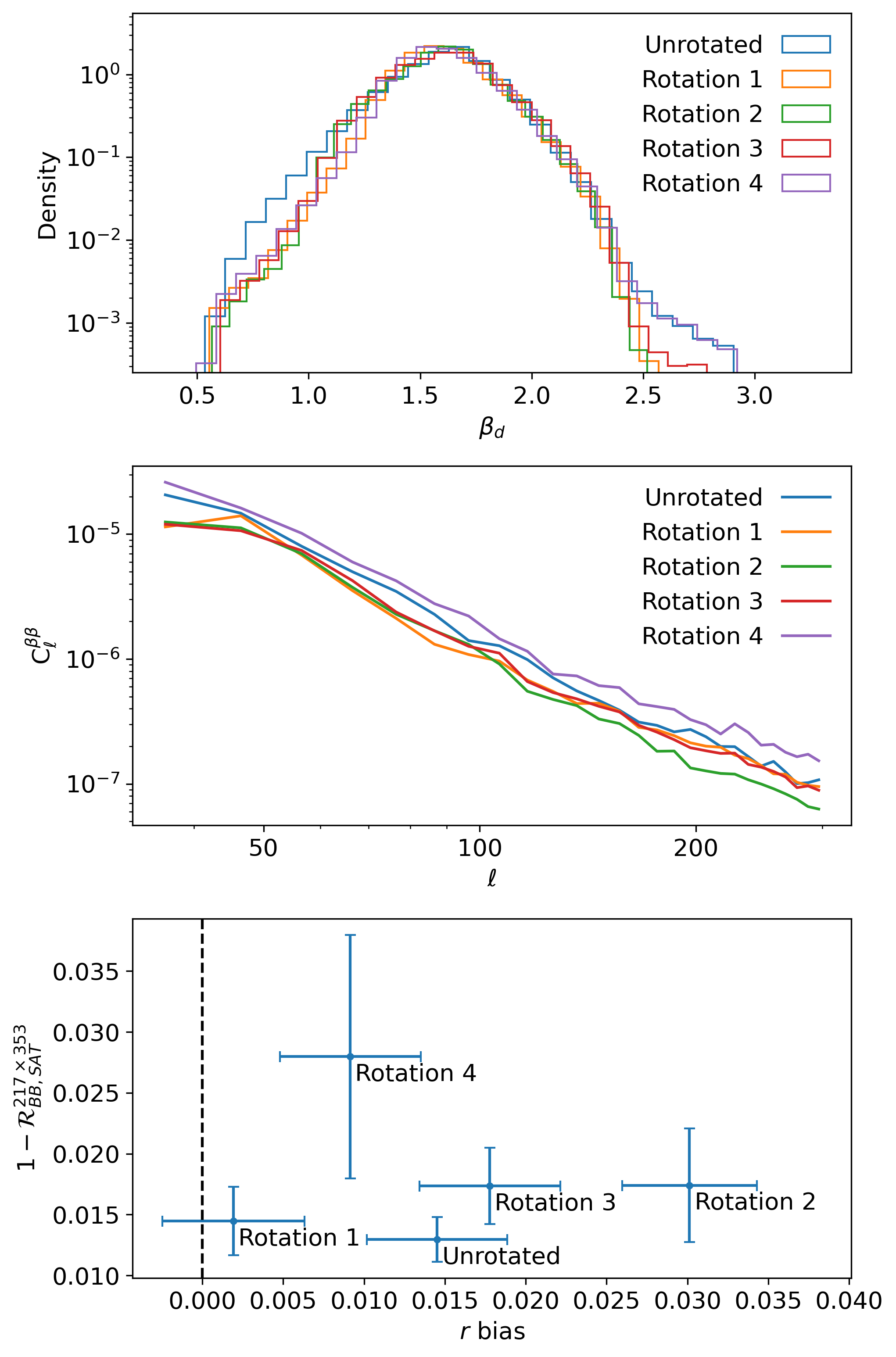}
    \caption{Selected properties of $\beta_d$ under the four different rotations considered. The distributions for 
    $\beta_d$ (top), and the corresponding power spectra (middle), in our considered SO region.  Both show small differences under different rotations, but without an obvious connection to the bias levels shown in Figure \ref{fig:d10_rot_r_bias}.
    Bottom: the decorrelation between the 217 and 353\,GHz frequencies in the SO SAT mask, and the corresponding bias in 
    $r$, for each rotation. There is not an obvious correlation between these two properties.}
    \label{fig:d10_rot_beta_prop}
\end{figure*}

To investigate this, we introduce the \textit{Ad10Bd10 rots} scenario described in Section~\ref{subsec:sim-SBd10}, where we apply a set of four arbitrary coordinate rotations to the $\beta_d$ map within the SO patch. These rotations effectively preserve the statistical properties (e.g., dispersion, angular power spectrum) of the $\beta_d$ field, while modifying any potential spatial or higher-order relationships with the dust amplitude map. We then propagate the rotated $\beta_d$ templates through the component separation pipeline and estimate $r$. These are shown in Figure~\ref{fig:d10_rot_r_bias}. 

\begin{figure*}[ht]
    \centering
    \includegraphics[width=0.7\linewidth]{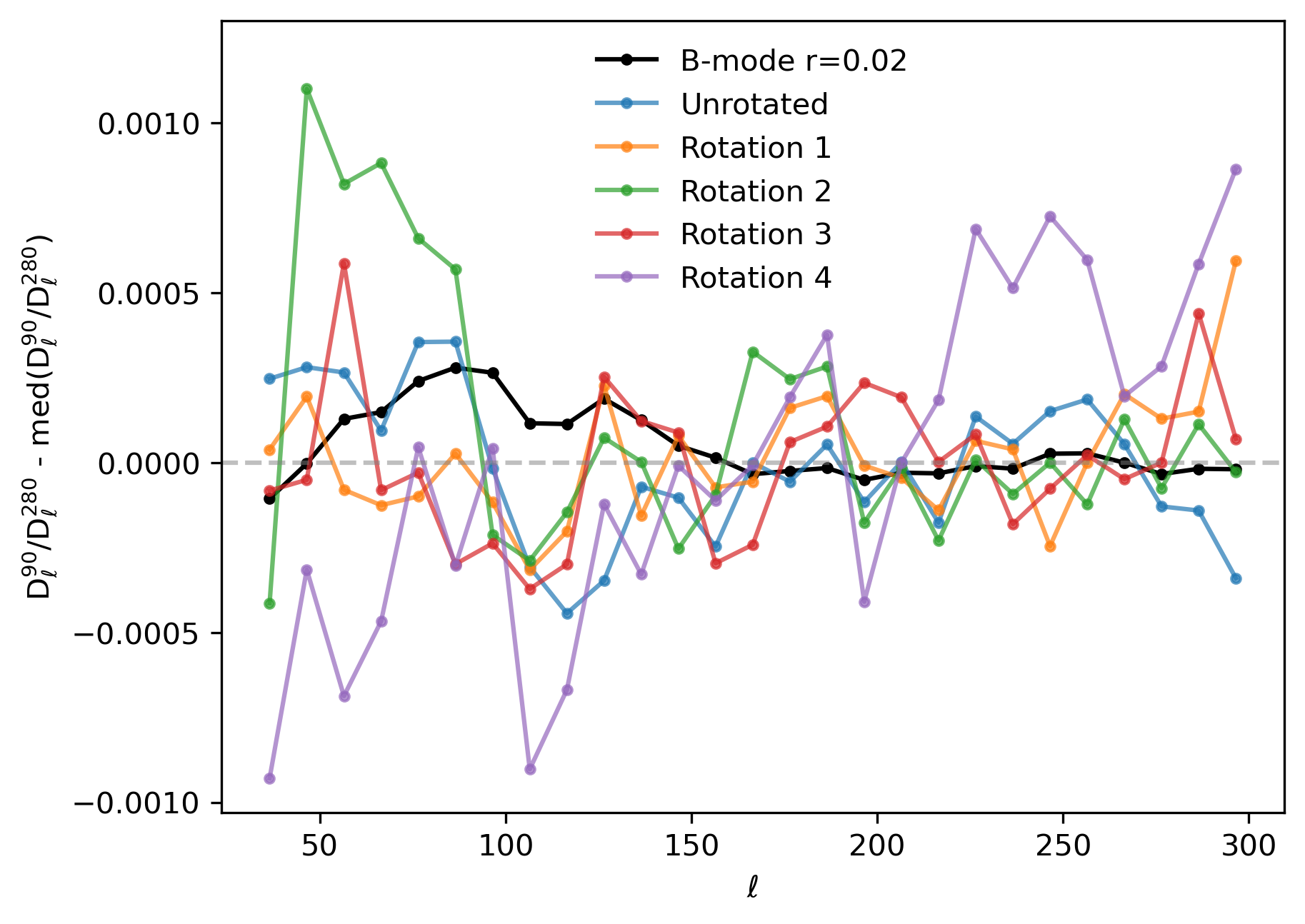}
    \caption{The ratios of the dust power spectra at 90~GHz to 280~GHz for the different $\beta_d$ rotations, with the median across bandpower bins subtracted.
    Variations across multipole reflects an $\ell$-dependent variation of the dust SED. 
    The black curve shows the ratio of primordial $B$-mode power spectrum at $r = 0.02$ to the standard \texttt{d10} 280\,GHz power spectrum. 
    Rotations that induce a larger bias in $r$ (e.g., Unrotated and Rotation 2) enhance the large-scale dust power at 90~GHz, mimicking tensor $B$-mode structure. This suggests that spatial features in $\beta_d$ are responsible for the bias in $r$.
    The residual structure in the median subtracted dust power spectral ratio qualitatively corresponds 
    to the excess large-scale power that the model absorbs as a larger---but incorrect---$r$ signal.}
    \label{fig:dust_dells_ratio}
\end{figure*}

Despite the fact that the rotated $\beta_d$ maps have nearly identical distributions and power spectra, both within each map and across different rotations as shown in Figure~\ref{fig:d10_rot_beta_prop}, we observe substantial variation in the $r$ distributions.
This indicates that neither the overall $\beta_d$ dispersion nor its spectral shape is responsible for the bias. 
Moreover, the bottom panel of Figure~\ref{fig:d10_rot_beta_prop} shows little correlation 
between frequency decorrelation and $r$ bias, further suggesting that the overall dispersion 
plays a limited role in driving the bias on $r$ in this scenario. Higher-order moments of the 
$\beta_d$ distributions, such as skewness and kurtosis, also exhibit no clear correlation with 
the estimated $r$ biases.
Instead, it appears to originate from the specific spatial structure of $\beta_d$ in each map, structure that is strongly spatially inhomogeneous, due to the physical distribution of dust on the sky.
 
Upon further investigation we find that certain configurations of the \texttt{d10} $\beta_d$ map, under specific rotations, modify the low-frequency dust angular power spectra in a way that resembles a $B$-mode signal. Figure~\ref{fig:dust_dells_ratio} shows the ratio of the 90~GHz and 280~GHz dust spectra under different $\beta_d$ rotations. While the 280~GHz spectra remain largely unaffected (given their proximity to the 353\,GHz pivot frequency), the 90~GHz spectra exhibit distinct $\ell$-dependent SED scaling under different $\beta_d$ rotations. 
Similar $\ell$-dependence is also observed in prior studies \citep[e.g.,][]{Chluba_2017_LOS}.
The differences in power spectral ratios are therefore dominated by the $\ell$-dependence of the 90~GHz dust power spectrum. In particular, some rotations, such as the standard \texttt{d10} and the Rotation 2, enhance large-scale ($\ell \lesssim 100$) power relative to small scales, producing a shape similar to the expected $B$-mode spectrum and resulting in a significant positive $r$ bias. Other rotations, such as Rotation 1, maintain flatter scaling and avoid introducing significant bias. 

\subsubsection{$\ell$-dependent SED effects}

To understand the $\ell$-dependent SED scaling of the dust power spectrum, we examine the power spectra of the moment-expansion dust maps. The $\beta_d$ template can be expressed as a perturbation from the mean $\bar{\beta}_d$:
\begin{align}
    \beta(\hat{n}) = \bar{\beta} + \delta\beta(\hat{n})
\end{align}
where $\hat{n}$ denotes the position in the sky.  The corresponding dust emission at frequency $\nu$ is given by the moment expansion:
\begin{align}
    \begin{aligned}
        m(\nu, \hat{n}) &= \left(\frac{\nu}{\nu_0}\right)^{\bar\beta + \delta\beta(\hat{n})}
        \frac{B(T, \nu)}{B(T, \nu_0)}m(\nu_0, \hat{n})\\
        &= \left[\left(\frac{\nu}{\nu_0}\right)^{\bar\beta}\frac{B(T, \nu)}{B(T, \nu_0)}m(\nu_0, \hat{n})\right]\\
        \Bigg[1 + \ln\left(\frac{\nu}{\nu_0}\right)&\delta\beta(\hat{n})
        + \frac{1}{2}\ln^2\left(\frac{\nu}{\nu_0}\right)\delta\beta(\hat{n})^2 + \O(\delta\beta^3)\Bigg]
    \end{aligned}
\end{align}
where $m(\nu_0, \hat{n})$ denotes the dust amplitude template at pivot frequency, and $B(T, \nu)$ is the Planck law at $T = 19.6$\,K. This moment expansion to the second order can be written as sum of three maps, or moment maps, that we call $m_0(\nu, \hat{n})$, $m_1(\nu, \hat{n})$, $m_2(\nu, \hat{n})$. These moment maps are powerful tools for interpreting foreground models \citep[e.g.,][]{Vacher_2024_how_bad_dust}.

\begin{figure*}[ht]
    \centering
    \includegraphics[width=\linewidth]{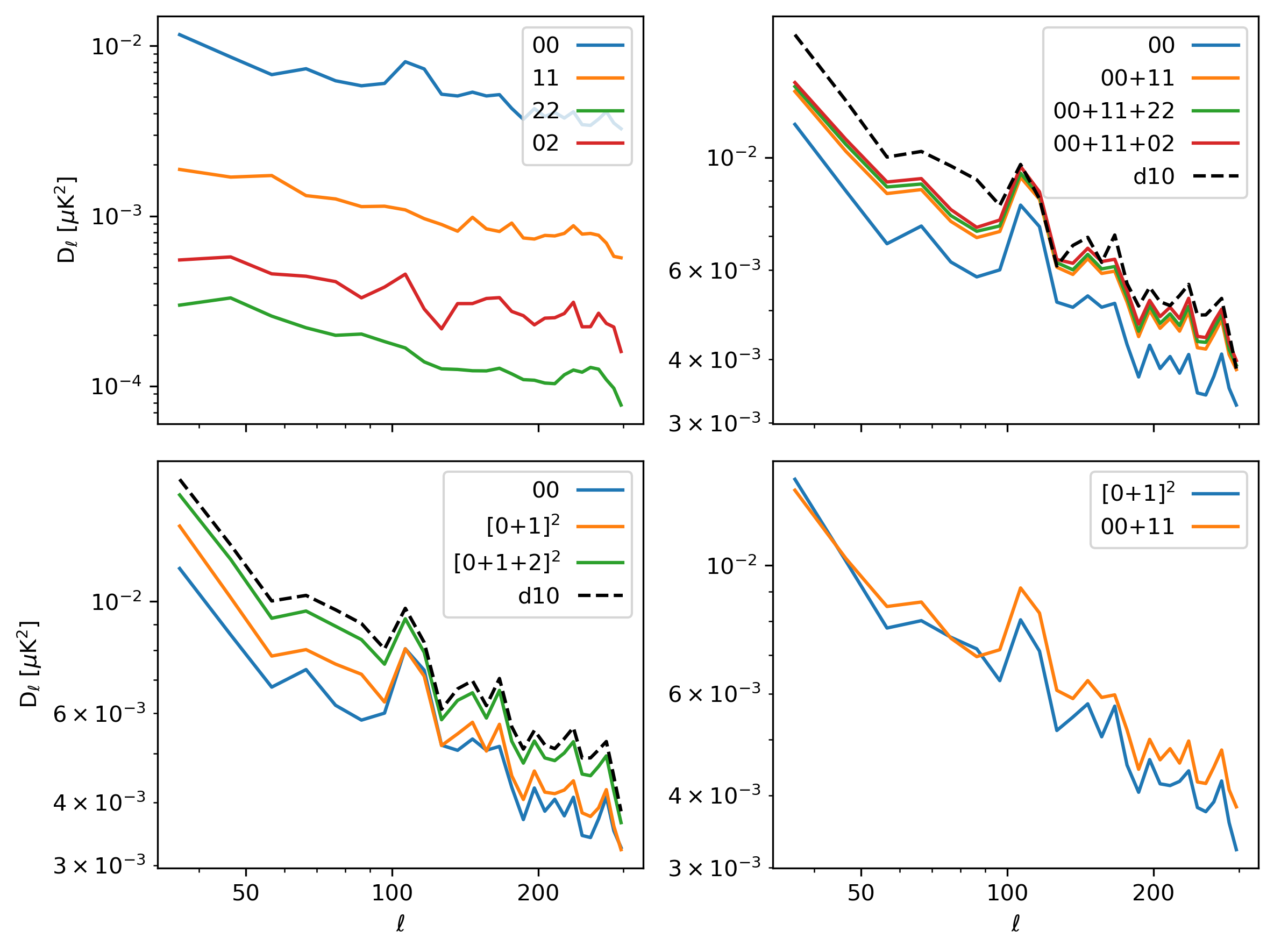}
    \caption{The 90~GHz power spectra of the leading-order moment maps in the \textit{$\beta_d$ spatial} scenario at $x_\beta = 1.6$. 
    \textbf{Top Left}: auto-spectra and selected cross-spectra of moment maps, indicated with a pair of numbers (e.g., 00 means 
    the auto spectra of the zeroth order moment map).
    \textbf{Top Right}: Selected combination of moment map auto-spectra and cross-spectra. The blue $00$ spectrum is used in our fiducial pipeline and clearly underestimates the \texttt{d10} spectrum (black). The red $00+11+02$ is the combination used in the moment method we consider in this analysis; it also does not fully capture the \texttt{d10} spectrum and explains why we see a biased result for $r$.
    \textbf{Bottom Left}: Selected power
    spectra for the sum of moment maps. We use square brackets to indicate power spectrum of map sums
    (e.g. $[0+1]^2$ means the power spectrum of the sum of zeroth and first order moment maps).
    The difference between the green line and black dashed line indicates the need for higher-order terms in this scenario where the dust is highly complex.
    \textbf{Bottom Right}: power spectra for the map sum and the sum of expansion power spectra to the first order. The difference between the two power spectrum shows the need to include the cross order term, 01, for accurate modeling.}
    \label{fig:d10_moment_ps}
\end{figure*}

Figure~\ref{fig:d10_moment_ps} illustrates the power spectra of these moment maps in the \textit{$\beta_d$ Spatial} scenario for $x_\beta = 1.6$, where the $\beta_d$ field remains unrotated. Since each $n$th-order moment map scales as $\ln^n\left(\frac{\nu}{\nu_0}\right)$, their contributions grow as the frequency moves away from $\nu_0$. This is especially significant at low frequencies, where higher-order terms begin to modify the dust power spectrum shape, leading to $\ell$-dependent SED effects.

The top-right panel of Figure~\ref{fig:d10_moment_ps} compares the power spectra for selected combinations of moment maps. The $00$, $00+11$, and $00+11+22$ curves correspond to zeroth-, second-, and fourth-order expansions (ignoring cross-moment correlations). While $m_0$ and $m_1$ contribute significantly, $m_2$ has negligible power. 

A particularly relevant combination is $00+11+02$, which corresponds to the existing second-order moment extension 
modeling in \citet{azzoni_2021_bbpower_moment} that we adopt in this work. This captures the dominant behavior but begins to deviate from the full \texttt{d10} power spectrum (black dashed line). This discrepancy is particularly relevant at larger scales, where if the model underestimates dust power, a positive bias in $r$ can be introduced as the pipeline compensates for the missing power. This deviation arises when either the level of frequency decorrelation is strong enough to induce high variance in $\delta\beta$, or the actual dust power spectrum deviates substantially from the assumed power-law form. Both effects are physically plausible, especially in realistic sky regions. 

In these cases, the current moment-based model may become underconstrained. Including 
higher-order terms, in addition to the minimal moment expansion implemented in 
\citet{azzoni_2021_bbpower_moment}, could better capture the observed behavior. 
Some examples can be found in \citet{Mangilli2021} and \citet{Vacher_2022_dust_moment}, where 
an extended moment expansion was applied in the contexts of Planck and \textit{LiteBIRD} component 
separation, respectively. However, doing so comes at a cost: each added term introduces 
additional parameters, which increases the uncertainty on $r$ (i.e., inflates $\sigma(r)$).

Another source of bias lies in the neglected cross-correlations between moment terms. As shown in the bottom-right panel of Figure~\ref{fig:d10_moment_ps}, the summed power of individual moment maps (orange line) differs from the power of their sum (blue line), indicating that cross-order correlations, especially between $m_0$ and $m_1$, contribute significantly to the total signal. These correlations may become non-negligible under strong frequency decorrelation. 

These results highlight the need to explore cross-order and higher-order moment-expansion terms for power spectral modeling.
As power spectral degeneracies become more challenging at higher sensitivity, modeling frameworks that incorporate both harmonic space and pixel space information may offer robust solutions if the true sky has complex spectral index variation \citep[e.g.,][]{Remazeilles_2021_fg_cmILC, azzoni_2023_hybrid_bbpower, Carones_2024_moment_model}. Importantly, such methods can enhance model fidelity without requiring additional nuisance parameters, thereby avoiding any potential degradation in $\sigma(r)$.

%% file: discussion.tex
\section{Discussion}
\label{sec:discussion}

In this work, we investigated the source of possible biases on the cosmological parameter $r$, arising from potential mismodeling of foreground spectral properties in component separation. Through a series of targeted simulations, including realistic spatially varying dust parameters and controlled covariance modeling, we identified conditions under which standard analysis pipelines can lead to biased $r$ estimates or degraded fit quality.

We quantified dust complexity in terms of frequency decorrelation, controlled through variations in 
the dispersion of SED parameters. We found a clear trend: increasing SED parameter dispersion leads to 
a stronger bias in $r$ when modeled with a power spectrum approach, with variations in $\beta_d$ having 
the largest effect. 
Section~\ref{subsec:s5d10-fg} noted that the anti-correlation between $\beta_d$ and $T_d$ templates can 
reduce the net decorrelation. Disentangling the origin of this anti-correlation is challenging for CMB experiments,
as independently constraining $T_d$ requires observations at higher frequencies typically in the THz range. 
In the absence of such information, it is difficult to determine whether the observed 
$\beta_d$-$T_d$ anti-correlation reflects physical dust properties or fitting degeneracies.

Focusing on $\beta_d$, we identified $\ell$-dependent SED scaling as a primary driver of mismodeling that leads to a bias in $r$. Through moment-expansion analysis of dust maps, we showed that higher-order moment terms alter the power spectrum shape, particularly in lower frequency channels.

To assess the adequacy of power spectrum modeling, we evaluated the goodness-of-fit using 
$\chi^2$ values. Models that assume a simple power-law angular power spectrum for foregrounds 
yield poor fits to the synchrotron simulations we considered. To improve fitting quality, modifications were needed in both the model definition 
and the covariance matrix. For foreground power spectra, adopting 
methods such as a per-$\ell$ fitting formalism may better 
account for deviations from a simple power law. 
For the data covariance, simulations should include both 
frequency decorrelation and an estimate of the $\ell$-dependence of foreground power spectra to ensure 
a proper characterization of cross-channel correlations. 
Accurately determining the level of frequency decorrelation is challenging in practice, 
which complicates the generation of Gaussian simulations with appropriate decorrelation level for 
a simulation-based covariance. A practical approach is to first quantify the relationship between 
recovered moment amplitudes $(B_d, B_s)$ and input decorrelation levels, begin with a simple 
Gaussian covariance, and iteratively introduce decorrelation effects based on the observed 
moment amplitude until convergence is achieved.
Alternatively, an analytic solution to the data covariance could be used.
These modifications will be important for accurately assessing the goodness-of-fit as power spectrum 
measurements become more precise.

Expanding the moment-expansion formalism to include cross-order and higher-order terms, as well as 
relaxing the assumption of a power-law $\beta_d$ angular power spectrum, are promising directions for 
improving dust modeling accuracy, if needed by the data. However, such refinement must be balanced against the resulting increase in $\sigma(r)$. Future work could also explore the application of existing hybrid map-$C_\ell$ component separation approaches to these more complex foreground models. Ancillary datasets at other wavelengths may also be useful for constraining dust behavior and informing model selection. 

In practice, we anticipate assessing the degree of complexity of the Simons Observatory data by checking the goodness of fit of the simplest models, and then increasing the complexity of the model to determine whether the estimate of $r$ and the foreground parameters, and the goodness of fit, changes when adding higher order terms. We would also expect to perform robustness tests to check the stability of our results derived from different subsets of the data (for example, excluding the 90~GHz or 150~GHz data, or using different regions of the sky), which could also reveal the need to enhance the model complexity. Implementing different foreground cleaning methods will continue to be valuable in assessing the consistency of results. 

%% file: conclusion.tex
\section{Conclusions}
\label{sec:conclusion}
Our analysis highlights several key findings regarding the origin of foreground-induced biases in tensor-to-scalar ratio measurements. We summarize here the principal conclusions of our work.

\begin{enumerate}

    \item Incorporating the effect of frequency decorrelation into the covariance matrix estimation was essential for achieving acceptable goodness of fit in the analyses presented here (see Figure~\ref{fig:gauss_r_chi2}). While beyond the scope of this work, full data-driven covariance estimation may be critical for future analyses on real data.

    \item Spatial variations in $\beta_d$ consistently lead to the strongest frequency decorrelation for a given scaling $x_{\beta_d}$ (Figure~\ref{fig:beta_d_scaling}) and the largest biases in the recovered tensor-to-scalar ratio $r$ (Figure~\ref{fig:realistic_r_result}).  For some of the higher values of the \texttt{d10} $\beta_d$ scaling explored ($ 1.2 \leq x_{\beta_d} \leq 1.5$ ), the induced decorrelation remains within the range allowed by the physically-motivated \texttt{d12} model, yet produces a non-negligible bias in $r$. This trend is observed in both Gaussian and more realistic foreground simulations, underscoring $\beta_d$ as the most influential parameter in driving frequency decorrelation and associated biases.
    
    \item When isolating the contributions of dust amplitude and $\beta_d$ (Figure~\ref{fig:mixture-result}), we find that the bias in $r$ is strongly correlated with the choice of $\beta_d$, while the amplitude map has little impact. This confirms that it is the non-Gaussian structure of $\beta_d$, and potentially its coupling with the dust amplitude, that drives the residual $r$ bias – especially in the high-dispersion regime where even the moment-expansion method fails to fully recover an unbiased estimate.
    
   \item  While other potential sources of bias, such as non-Gaussianity in the dust amplitude, limitations of the power-law spectral model, or lowest-order correlations between amplitude and $\beta_d$, cannot be fully excluded, they are unlikely to be dominant. This is supported by (i) the equivalence in decorrelation between Gaussian and non-Gaussian $\beta_d$ fields, (ii) the good PtE values obtained when using a power-law dust model, and (iii) explicit tests showing that low-order correlations between $\beta_d$ and amplitude (Figure~\ref{fig:analytic_correlation}) have negligible impact on the bias in $r$ (Figure~\ref{fig:gauss_r_chi2}). These results point to missing higher-order statistics in the current modeling framework as the most plausible origin of the residual bias.
    
    \item The rotation analysis (Figure~\ref{fig:d10_rot_r_bias}), in which the $\beta_d$ map is rotated independently of the amplitude, reveals sensitivity to the spatial coupling between $\beta_d$ and the dust amplitude. This supports the interpretation that a bias on $r$ can arise from non-trivial correlations between these fields. Figure~\ref{fig:dust_dells_ratio} further shows that different $\beta_d$ maps yield distinct $\ell$-dependence in the dust power spectrum, reinforcing the connection between the bias and changes in spectral shape across frequency.

    \item While non-Gaussian amplitudes have a modest impact on $r$ biases relative to Gaussian amplitudes with the same power spectrum, the non-Gaussianity of the $\beta$ map has a significant effect. 
    Figure~\ref{fig:d10_moment_ps} indicates that our current moment expansion (including the terms $00+11+02$) underfits the decorrelation observed in the re-scaled \texttt{d10} simulation (with $x_{\beta_d}=1.6$), pointing to missing higher-order moment terms in the model. These terms include (i) the ones dependent on the $\beta_d$-amplitude correlations, and (ii) the moment terms of order higher than 2, which quantify non-Gaussianities of the $\beta_d$ distribution \citep{Vacher_2024_how_bad_dust}.
\end{enumerate}

  Future extensions of the analysis methods employed here could include higher-order terms in the expansion, which are straightforward to include in the current implementation of the moments method. Although this may increase parameter degeneracy and increase $\sigma(r)$, it would provide a more complete description of foreground complexity if the data require it. Alternative approaches could include incorporating the pixel-level correlations in the current model, as in the ``hybrid''  method proposed in \citet{azzoni_2023_hybrid_bbpower}, 
  or leverage moments-based frameworks operating in other domains beyond the power spectrum \citep[e.g.,][]{Remazeilles_2021_fg_cmILC, Carones_2024_moment_model}.
  Improved measurements will enable sharper foreground characterization, guiding the development of increasingly robust modeling techniques.